\documentclass[lettersize,journal]{IEEEtran}
\usepackage{amsmath,amsfonts}
\usepackage{algorithmic}
\usepackage{algorithm}
\usepackage{array}
\usepackage[caption=false,font=normalsize,labelfont=sf,textfont=sf]{subfig}
\usepackage{textcomp}
\usepackage{stfloats}
\usepackage{url}
\usepackage{verbatim}
\usepackage{graphicx}
\usepackage{cite}
\usepackage{cuted}
\usepackage{upgreek}
\usepackage{bm}
\usepackage{amssymb}
\usepackage{xcolor}
\usepackage{mathtools}
\usepackage{multirow}
\allowdisplaybreaks 

\begin{document}
\bstctlcite{IEEEexample:BSTcontrol}

\title{A physics-guided data-driven feedforward tracking controller for systems with unmodeled dynamics -- applied to 3D printing}

\author{Cheng-Hao Chou, Molong Duan, Chinedum E. Okwudire

\thanks{This paper is supported by National Science Foundation grant \#1931950 and a grant from CISCO Systems Inc.}

\thanks{C.-H. Chou (email: cchengha@umich.edu) and C.E. Okwudire (email: okwudire@umich.edu) are with the Department of Mechanical Engineering, University of Michigan, Ann Arbor, MI 48109 USA. M. Duan (e-mail: duan@ust.hk) is with the Department of Mechanical and Aerospace Engineering, Hong Kong University of Science and Technology, Hong Kong SAR, China.}}

\IEEEpubid{This work has been submitted to the IEEE for possible publication. Copyright may be transferred without notice, after which this version may no longer be accessible.}

\maketitle

\begin{abstract}
A hybrid (i.e., physics-guided data-driven) feedforward tracking controller is proposed for systems with unmodeled linear or nonlinear dynamics. The controller is based on the filtered basis function (FBF) approach, hence it is called a hybrid FBF controller. It formulates the feedforward control input to a system as a linear combination of a set of basis functions whose coefficients are selected to minimize tracking errors. The basis functions are filtered using a combination of two linear models to predict and minimize the tracking errors. The first model is physics-based and remains unaltered during the execution of the controller, while the second is data-driven and is continuously updated during the execution of the controller. To ensure its practicality and safe learning, the proposed hybrid FBF controller is equipped with the ability to handle delays in data acquisition and to detect impending instability due to its inherent data-driven feedback loop. Its effectiveness is demonstrated via application to vibration compensation of a 3D printer with unmodeled linear and nonlinear dynamics. Thanks to the proposed hybrid FBF controller, the tracking accuracy of the 3D printer is significantly improved in experiments involving high-speed printing, compared to a standard FBF controller that does not incorporate a data-driven model. Furthermore, the ability of the hybrid FBF controller to detect and, hence, potentially avoid impending instability is demonstrated offline using data collected online from experiments. 
\end{abstract}


\section{Introduction} \label{introduction}
\IEEEPARstart{T}{racking} control is important in a wide range of applications, such as manufacturing, robotics, and aeronautics. The goal of the tracking control is to force the output of a dynamic system to follow a desired trajectory by minimizing the tracking error, i.e., the difference between the output and the desired trajectories. In particular, feedforward (FF) techniques play an important role in tracking control. Compared to feedback (FB) control, FF control can preemptively compensate tracking error, which is impossible using FB control. In some applications, like 3D printing, where open-loop-controlled stepper motors are typically used for motion generation, FF is the only resource for control. Moreover, even in applications where sensors are available, they may be unsuitable for real-time FB control due to practical constraints, like sensor and data acquisition delays, excessive noise, observability or stability concerns. Moreover, FF control can always be combined with FB control to improve overall tracking accuracy.

Theoretically, perfect FF tracking control can be achieved by exact model inversion. In linear systems, the model inversion can be simplified to pole-zero cancellation. However, due to the common occurrences of uncancellable zeros and uncertainty in system dynamics, exact model inversion may lead to unbounded control signals or poor tracking performance in practice. To solve this issue, various FF controllers based on approximate model inversion have been proposed in the literature, as summarized in review papers like \cite{clayton2009review, rigney2009nonminimum, van2018inversion}. Examples include the zero phase error tracking controller (ZPETC) \cite{tomizuka1987zero}, zero magnitude error tracking controller (ZMETC) \cite{butterworth2012analysis}, extended bandwidth ZPETC \cite{torfs1992extended}, and model matching controller \cite{wen2004experimental}, to name a few. A relatively recent addition to the available tracking control methods is the filtered basis function (FBF) approach \cite{ramani2017tracking, duan2018limited, ramani2019optimal, romagnoli2019general, ramani2021optimal, ramani2020robust}. The origin of the FBF approach can be traced back to the work of Frueh and Phan \cite{frueh2000linear} on inverse linear quadratic iterative learning control (ILC). It formulates the FF control input to a system as a linear combination of a set of basis functions that are filtered using the system’s model to predict the system's output. Thus, it allows the unknown coefficients of the basis functions to be selected to minimize tracking error. A major advantage of the FBF approach over several other FF tracking control techniques is its versatility; it is applicable to any linear system and has been shown to yield accurate tracking regardless of the location of zeros of the system in the complex plane \cite{ramani2017tracking}. One application that has benefited from the versatility of the FBF approach is tracking control of vibration-prone 3D printers, which have a variety of underlying dynamics \cite{duan2018limited, ramani2021optimal, ramani2020robust, edoimioya2021software}.

\IEEEpubidadjcol

A major deficiency of standard FF tracking control techniques is their inability to handle unmodeled dynamics prevalent in practice. For example, the vibration dynamics of 3D printers often have unmodeled nonlinear dynamics, e.g., due to friction, backlash and nonlinear belt stiffness \cite{duan2018limited, ramani2021optimal}. These nonlinear dynamics are left out of linear models used for FF tracking control. Moreover, the vibration dynamics of 3D printers vary over time due to printer wear or end-user modifications. These changes to the printer dynamics are unknown, hence not captured in the models used for FF tracking control. A variety of strategies have been proposed in the literature to enable FF tracking controllers to handle unmodeled dynamics. They can be broken down into two main categories: (1) iterative learning control (ILC) and (2) adaptive FF control. ILC updates feedforward control input signals based on data gathered from past iterations \cite{bristow2006survey}. Hence, it can learn and adapt to unmodeled dynamic behaviors. A major advantage of ILC is that it does not require a parametric model to operate, hence it can learn unmodeled dynamics with an unknown structure. However, ILC's assumption of repeating trajectories or disturbances do not apply to many practical scenarios, such as tracking control in 3D printing where motions are typically nonrepetitive due to customization. Moreover, when the stability or convergence is being analyzed, the broad knowledge of plant dynamics is typically required. Adaptive FF controllers, on the other hand, tune the parameters of the controllers to account for unmodeled or changing dynamics \cite{boeren2015iterative, dumanli2020data, tsao1987adaptive, huang2021data}. While there are some adaptive FF controllers that must be tuned using repeating trajectories, e.g., \cite{boeren2015iterative, dumanli2020data}, it is not always a requirement. However, a shortcoming of adaptive FF controllers is that they typically require a fixed structure with associated parameters to be tuned online \cite{boeren2015iterative, dumanli2020data, tsao1987adaptive, huang2021data}, despite the fact that the stability properties can be obtained more easily due to fixed structure. This diminishes their usefulness in situations where the exact structure of the unmodeled dynamics is a priori unknown or changing, such as the unknown aftermarket modification of 3D printers.

To handle both nonrepeating trajectories and the unmodeled dynamics with unknown structure, neural networks (NN) have also been incorporated into FF controllers, due to their loose structure and ability to approximate any mathematical function \cite{Kon2022PhysicsGuidedNN, otten1997linear, bolderman2022feedforward, ishihara2009protection, pan2017biomimetic}. However, their high dimensional and nonlinear nature make them difficult to be trained online or their stability to be rigorously analyzed to ensure safe learning and control. As a result, most existing works on NN-based FF controllers either ignore the stability of the NN \cite{Kon2022PhysicsGuidedNN, otten1997linear, bolderman2022feedforward} or analyze its stability under very strict conditions that are not validated experimentally \cite{ishihara2009protection, pan2017biomimetic}.

As an alternative approach, the present authors have proposed a hybrid (i.e., physics-guided data-driven) model for servo systems entirely based on linear models \cite{chou2021linear}. In the hybrid model, the output of a structured linear physics-based model is fed into a loosely structured linear data-driven model that is continuously updated online. It was shown through simulations and experiments that, thanks to the loose structure of its data-driven component, the hybrid model was able to learn unknown nonlinear dynamics and improve overall model prediction compared to its purely physics-based counterpart. A preliminary version \cite{chou2021hybrid} of the present paper has incorporated the linear hybrid model into the FBF approach, where it was shown in simulations that the resulting hybrid FBF controller could significantly improve the tracking performance of a system with unmodeled nonlinear dynamics. However, the preliminary investigation failed to address two issues that are key to the practicality of the hybrid FBF controller, namely: (1) how to handle delays in acquiring data for training the data-driven model, and (2) how to assess the stability of the hybrid FBF controller due to the inherent feedback loop introduced by its data-driven component. Moreover, the preliminary work did not validate the effectiveness of the hybrid FBF controller in experiments. To address these shortcomings, together with its preliminary version \cite{chou2021hybrid}, the original contributions of this paper are:
\begin{enumerate}
\item{In Section~\ref{method}, proposing a physics-guided data-driven FF tracking controller based on the FBF approach. The proposed method, called the hybrid FBF controller, is able to provide accurate tracking for systems with unmodeled dynamics. Moreover, it is formulated to account for measurement delays which occur in practice.}
\item{In Section~\ref{stability}, proposing a rigorous stability analysis approach that allows impending instability of the hybrid FBF controller to be detected in order to avert instability.}
\item{In Section~\ref{experiments}, demonstrating the effectiveness of the proposed hybrid FBF controller and stability analysis methods in online and offline experiments on a vibration-prone 3D printer with unmodeled linear and nonlinear dynamics.}
\end{enumerate}
These contributions are followed by conclusions and the future work in Section \ref{conclusion}.

\section{Formulation of the Hybrid FBF Controller} \label{method}
\subsection{Review of the Linear Hybrid Model} \label{LHM}

\begin{figure}[!t]
\centering
\includegraphics[width=\columnwidth]{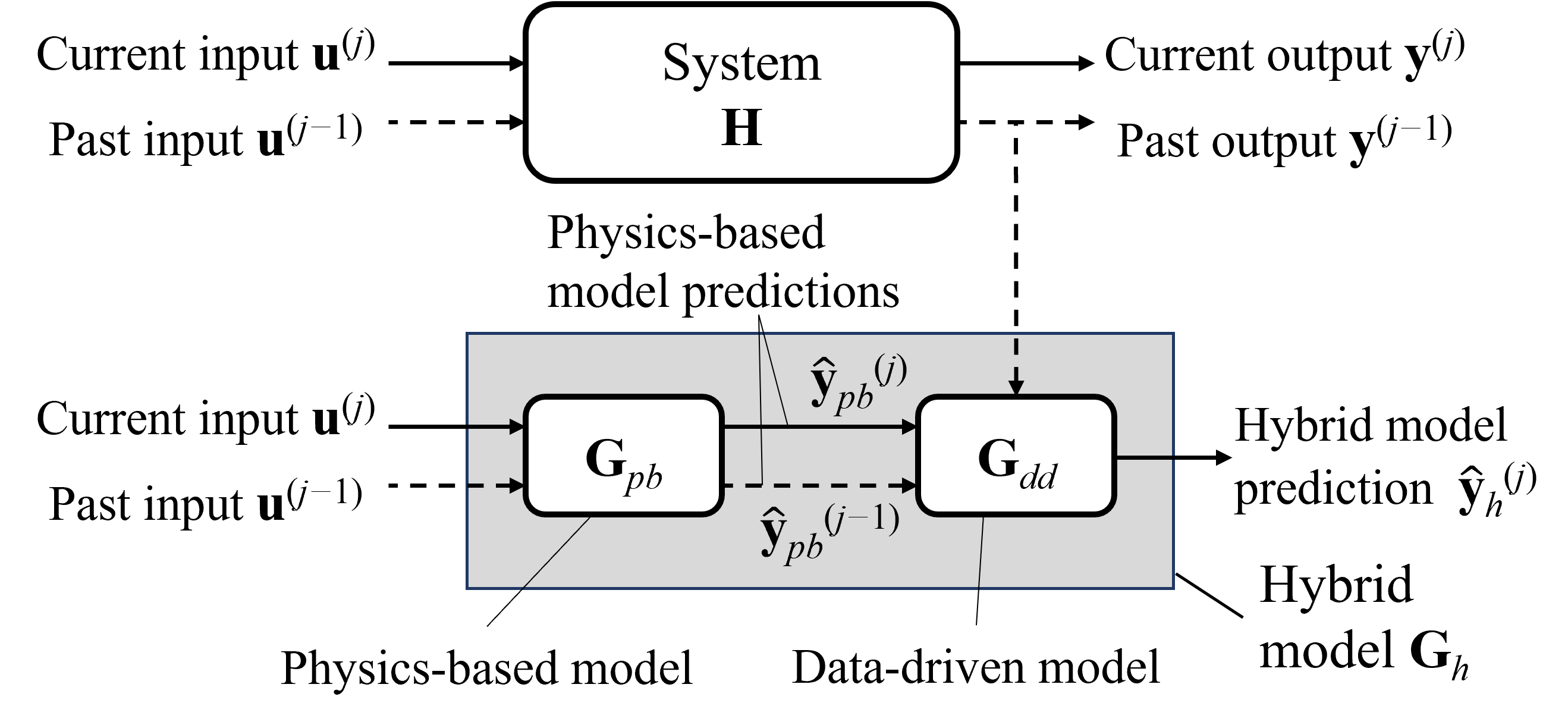}
\caption{Overall framework of the linear hybrid model $\mathbf{G}_h$ \cite{chou2021linear}.}
\label{lhm_fig}
\end{figure}

\begin{figure}[!t]
\centering
\includegraphics[width=0.8\columnwidth]{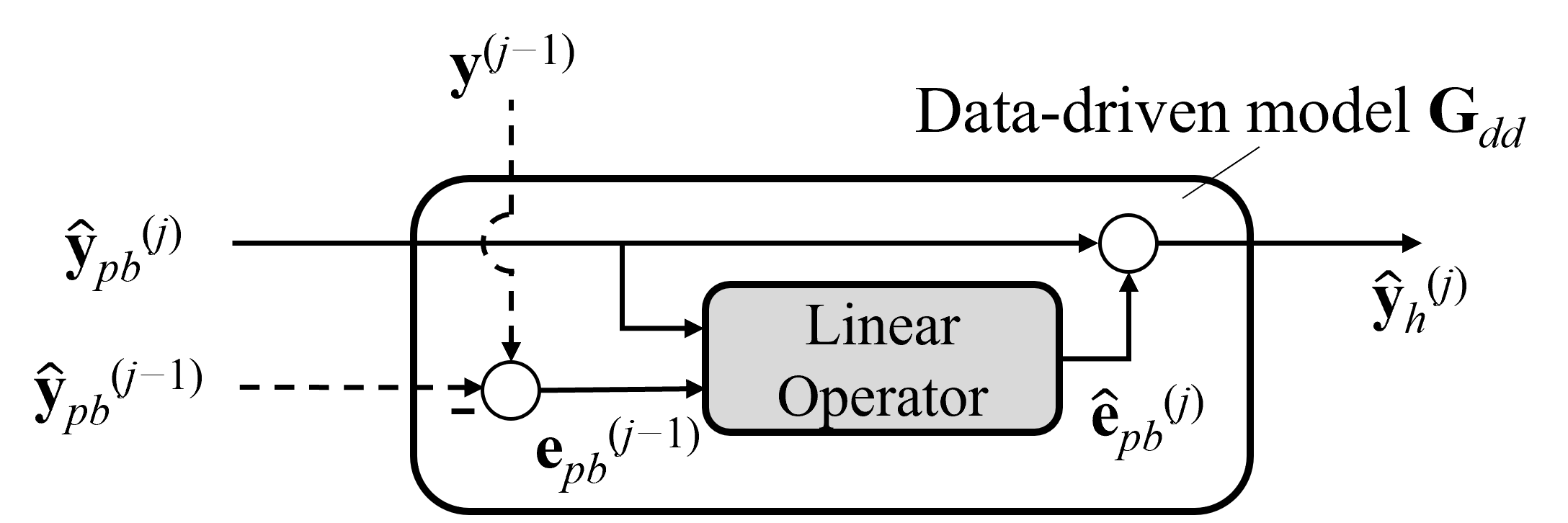}
\caption{Detailed mechanism of the data-driven portion of $\mathbf{G}_h$ \cite{chou2021linear}.}
\label{lhm_dd}
\end{figure}

Consider a stable and causal SISO dynamic system $\mathbf{H}$, which is sampled at a constant interval $T_s$. The input and output of $\mathbf{H}$ are respectively denoted by $u(k)$ and $y(k)$, where $k=0,1,2,\ldots$, is the time step sampled at the fixed interval $T_s$. In addition, suppose that the inputs and outputs are generated and measured in small batches. The signal in the $j$-th batch is denoted by the superscript $(j)$ and is defined, taking the measured output $y$ as an example, as the vector  
\begin{equation} \label{def_batch}
    \mathbf{y}^{(j)} \triangleq \begin{bmatrix} y(jN),\; y(jN+1),\; \ldots,\; y((j+1)N-1) \end{bmatrix}^\top
\end{equation}
where $N$ is the length of the small batches and $j=0,1,2,\ldots$ is the batch index.

For traditional FF tracking controllers, a physics-based model $\mathbf{G}_{pb}$, such as a transfer function or a lifted representation, is used to predict the response of $\mathbf{H}$. The controller then drives its prediction, denoted as $\mathbf{\hat{y}}_{pb}$, to be as close as possible to the desired trajectory $\mathbf{y}_d$. However, when there exists unmodeled dynamics that are not captured by $\mathbf{G}_{pb}$, the inaccurate prediction of $\mathbf{\hat{y}}_{pb}$ can degrade the performance of the controller. 

To enhance the prediction accuracy of servo systems, the authors have proposed a linear hybrid model \cite{chou2021linear}. As shown in Fig.~\ref{lhm_fig}, the linear hybrid model $\mathbf{G}_h$ cascades a linear physics-based model $\mathbf{G}_{pb}$ and a linear data-driven model $\mathbf{G}_{dd}$  to output a more accurate prediction $\mathbf{\hat{y}}_h$. More specifically, $\mathbf{G}_{dd}$ aims to complement $\mathbf{G}_{pb}$ using past measured data. To predict the output in the $j$-th batch, it takes the past and current physics-based predictions, $\mathbf{\hat{y}}_{pb}^{(j-1)}$ and $\mathbf{\hat{y}}_{pb}^{(j)}$, as well as the past measured output $\mathbf{y}^{(j-1)}$ as inputs, i.e., 
\begin{equation} \label{lhm_general}
    \begin{split}
    \mathbf{\hat{y}}_{h}^{(j)} 
    & = \mathbf{G}_{dd} \left(\mathbf{\hat{y}}_{pb}^{(j-1)}, \mathbf{\hat{y}}_{pb}^{(j)}, \mathbf{y}^{(j-1)} \right) \\
    & = \mathbf{G}_{dd} \left(\mathbf{G}_{pb}\mathbf{u}^{(j-1)}, \mathbf{G}_{pb}\mathbf{u}^{(j)}, \mathbf{y}^{(j-1)} \right) \\
    & = \mathbf{G}_{h} \left(\mathbf{u}^{(j-1)}, \mathbf{u}^{(j)}, \mathbf{y}^{(j-1)} \right)
    \end{split}
\end{equation}

In more detail, $\mathbf{G}_{dd}$ is constructed based on a linear regression model, and its detailed structure is shown in Fig.~\ref{lhm_dd}. The linear regression model estimates the physics-based model prediction error $\mathbf{e}_{pb} \triangleq \mathbf{y}-\mathbf{\hat{y}}_{pb}$ recursively for each time step, which is then added to $\mathbf{\hat{y}}_{pb}$ to obtain the final hybrid model prediction $\mathbf{\hat{y}}_{h}$. Therefore, using $\mathbf{G}_h$, $\mathbf{\hat{y}}_h^{(j)}$ is computed by recursively applying
\begin{equation} \label{lhm_recursive}
    \hat{y}_{h}(k) = \hat{y}_{pb}(k) + \hat{e}_{pb}(k) = \hat{y}_{pb}(k) + {{}\mathbf{\hat{w}}^{(j)}}^\top \bm{\upphi}(k)
\end{equation}

\noindent for each time step $k$ belongs to the the $j$-th batch, i.e., $k \in \{ jN,\; jN+1,\; \ldots,\; (j+1)N-1 \}$, where $\bm{\upphi}(k)$ is the feature vector for the prediction of $e_{pb}(k)$ and $\mathbf{\hat{w}}^{(j)}$ is the model weights trained with the data prior to the $j$-th batch. Furthermore, the feature vector $\bm{\upphi}(k)$ is designed as
\begin{equation} \label{lhm_featurevec}
    \begin{split}
    \bm{\upphi}(k) = \big[ 1, \quad & \hat{y}_{pb}(k-q+1), \quad \ldots, \quad \hat{y}_{pb}(k), \\
    & e_{pb}(k-p), \quad \ldots, \quad e_{pb}(k-1) \big]^\top
    \end{split}
\end{equation}

\noindent where $q$ and $p$ are design parameters representing the number of ${\hat{y}_{pb}}$ and ${e}_{pb}$ terms included in $\bm{\upphi}$. Besides, for all unavailable ${e}_{pb}$, i.e., ${e}_{pb}(k)$ for $k \geq jN$, ${e}_{pb}$ is replaced with the estimated values $\hat{e}_{pb}$ defined in Eq.~\eqref{lhm_recursive}. On the other hand, $\mathbf{\hat{w}}^{(j)}$ is trained by minimizing the difference between ${e}_{pb}$ and $\hat{e}_{pb}$ for all time steps prior to the $j$-th batch, through the following regularized least squares optimization problem
\begin{equation} \label{lhm_train}
    \mathbf{\hat{w}}^{(j)}
    = \underset{\mathbf{w}}{\mbox{argmin}} \sum_{k=0}^{jN-1} [e_{pb}(k)-\mathbf{w}^\top \bm{\upphi}(k)]^2+\lambda \lVert \mathbf{w} \rVert_2^2
\end{equation}
where $\lambda > 0$ is a tunable regularization factor that prevents overfitting. 

\noindent {\bf{Remark 1:}} Even though the hybrid model $\mathbf{G}_h$ is linear, it is able to predict unmodeled nonlinear behavior by approximating it using the past measured data in a piecewise linear fashion \cite{chou2021linear}. Moreover, though $\mathbf{G}_h$ is loosely parameterized with $q$ and $p$, it regresses on past prediction errors caused by unmodeled dynamics, thus making it versatile in its representation of unmodeled dynamics using linear combinations of past errors \cite{chou2021linear}.

\noindent {\bf{Remark 2:}} The recursive nature of $\mathbf{G}_h$ described in Eq.~\eqref{lhm_recursive} allows it to predict a system's output with a prediction horizon of arbitrary length. However, in practice, its prediction accuracy drops as the horizon expands.

The hybrid model prediction of \mbox{$(j+1)$-th} batch can be formulated as 
\begin{equation} \label{lhm_extend}
    \begin{split}
    \mathbf{\hat{y}}_{h}^{(j+1)} 
    & = \mathbf{G}_{dd} \left(\mathbf{\hat{y}}_{pb}^{(j)}, \mathbf{\hat{y}}_{pb}^{(j+1)}, \mathbf{\hat{y}}_{h}^{(j)} \right) \\
    & = \mathbf{G}_{dd} \left(\mathbf{\hat{y}}_{pb}^{(j)}, 
                        \mathbf{\hat{y}}_{pb}^{(j+1)}, 
                        \mathbf{G}_{dd}(\mathbf{\hat{y}}_{pb}^{(j-1)}, \mathbf{\hat{y}}_{pb}^{(j)}, \mathbf{y}^{(j-1)})
                        \right) \\
    & = \mathbf{G}_{dd,2} \left(\begin{bmatrix} \mathbf{\hat{y}}_{pb}^{(j-1)} \\ \mathbf{\hat{y}}_{pb}^{(j)} \end{bmatrix}, 
                          \mathbf{\hat{y}}_{pb}^{(j+1)}, 
                          \mathbf{y}^{(j-1)}
                          \right) \\
    & = \mathbf{G}_{h,2} \left(\begin{bmatrix} \mathbf{u}^{(j-1)} \\ \mathbf{u}^{(j)} \end{bmatrix}, 
                          \mathbf{u}^{(j+1)}, 
                          \mathbf{y}^{(j-1)}
                          \right) \\
    \end{split}
\end{equation}
where $\mathbf{G}_{dd,2}$ and $\mathbf{G}_{h,2}$ respectively define the data-driven and hybrid prediction models for two batches ahead.

\subsection{Derivation of Hybrid FBF Controller} \label{HybridFBF}

\begin{figure*}[!t]
\centering
\includegraphics[width=0.8\textwidth]{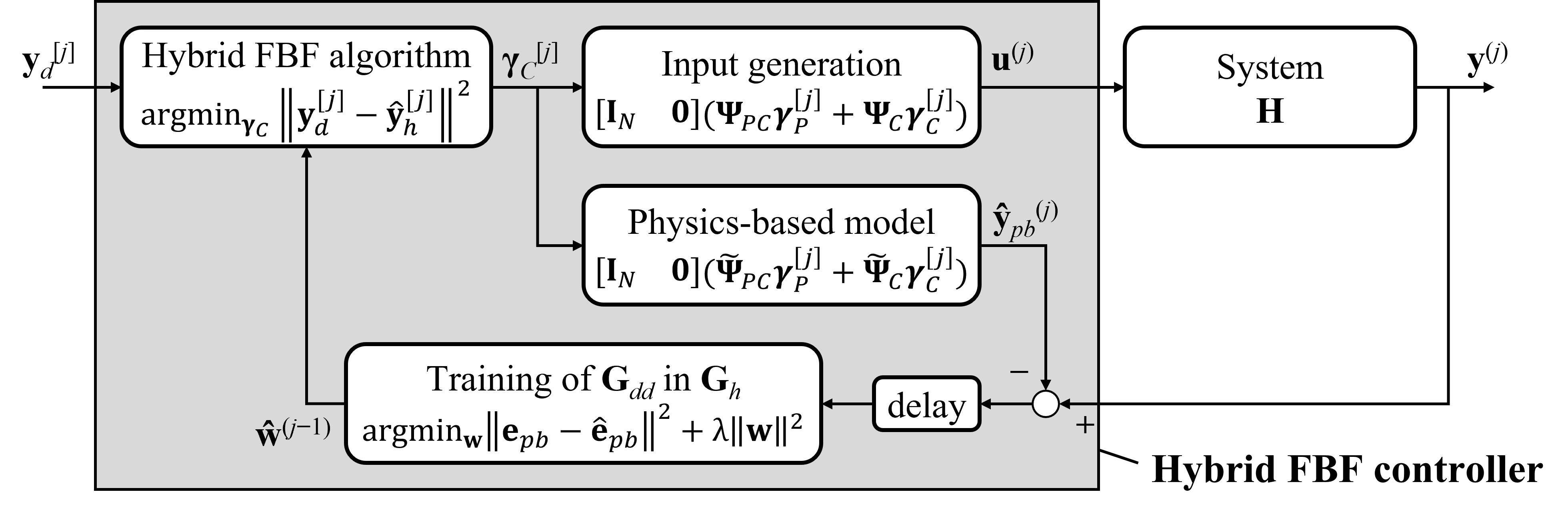}
\caption{Simplified diagram for the overall framework of the hybrid FBF approach.}
\label{hybridfbf}
\end{figure*}

As described in Sections~\ref{introduction} and \ref{LHM}, the task of a FF tracking controller is to minimize the error between the predicted system response $\mathbf{\hat{y}}$ and the desired trajectory $\mathbf{y}_d$. Since the performance of the FF controller is closely linked to the accuracy of its prediction model, here, we seek to apply the linear hybrid model $\mathbf{G}_h$ to the receding horizon version of the standard FBF approach \cite{duan2018limited} to enhance its tracking performance.

Since $\mathbf{G}_h$ is linear, according to Eq.~\eqref{lhm_general} and Eq.~\eqref{lhm_recursive}, the hybrid model prediction $\mathbf{\hat{y}}^{(j)}$ can be expressed as linear operation involving ${\mathbf{\hat{y}}_{pb}^{(j-1)}}$, ${\mathbf{\hat{y}}_{pb}^{(j)}}$, and ${\mathbf{e}_{pb}^{(j-1)}}$ as
\begin{equation} \label{lhm_largematrix}
    \mathbf{\hat{y}}_{h}^{(j)} 
    = \mathbf{G}_{dd} \left( \mathbf{\hat{y}}_{pb}^{(j-1)}, \mathbf{\hat{y}}_{pb}^{(j)},\mathbf{y}^{(j-1)} \right)
    = \mathbf{L}_{dd}^{(j)} \cdot 
        \begin{bmatrix} 
        1 \\
        {\mathbf{\hat{y}}_{pb}^{(j-1)}} \\
        {\mathbf{\hat{y}}_{pb}^{(j)}} \\
        {\mathbf{e}_{pb}^{(j-1)}} \\
        \end{bmatrix}
\end{equation}

\noindent where the $\mathbf{L}_{dd}^{(j)}$ matrix is constructed based on $\mathbf{\hat{w}}^{(j)}$ \cite{chou2021hybrid} and Eq.~\eqref{lhm_largematrix} can be rewritten as follows
\begin{equation} \label{lhm_matrixform}
    \mathbf{\hat{y}}_{h}^{(j)} =
    \mathbf{L}_{dd,a}^{(j)} \mathbf{\hat{y}}_{pb}^{(j)} +
    \mathbf{L}_{dd,uy}^{(j)} \mathbf{\hat{y}}_{pb}^{(j-1)} +
    \mathbf{L}_{dd,ue}^{(j)} \mathbf{e}_{pb}^{(j-1)} +
    \mathbf{L}_{dd,u1}^{(j)}
\end{equation}
where the matrices $\mathbf{L}_{dd,a}^{(j)}$, $\mathbf{L}_{dd,uy}^{(j)}$, $\mathbf{L}_{dd,ue}^{(j)}$, and $\mathbf{L}_{dd,u1}^{(j)}$ are the submatrices of $\mathbf{L}_{dd}^{(j)}$.

Although the prediction result in Eq.~\eqref{lhm_matrixform} may generate the control input for the $j$-th batch, in practice, there are bound to be delays between the end of batch $j-1$ and the availability of the data captured at batch $j-1$. Therefore, $\mathbf{y}^{(j-1)}$ (and $\mathbf{e}_{pb}^{(j-1)}$) are not immediately available at the end of batch $j-1$, making the use of Eq.~\eqref{lhm_matrixform} impractical. To address this problem without loss of generality, suppose the delay in acquiring $\mathbf{y}^{(j-1)}$ is one batch length, i.e., the latest available measurement is $\mathbf{y}^{(j-2)}$ when optimizing the inputs for the $j$-th batch. As described in Eq.~\eqref{lhm_extend}, $\mathbf{y}^{(j-1)}$ can be substituted with $\mathbf{\hat{y}}_h^{(j-1)}$. In other words, the hybrid prediction $\mathbf{\hat{y}}_{h}^{(j)}$ can be rewritten as
\begin{equation} \label{lhm_delay}
    \mathbf{\hat{y}}_{h}^{(j)} =  
    \mathbf{G}_{dd,2}   
    \left(
    \begin{bmatrix} \mathbf{\hat{y}}_{pb}^{(j-2)} \\ \mathbf{\hat{y}}_{pb}^{(j-1)} \end{bmatrix}, 
    \mathbf{\hat{y}}_{pb}^{(j)}, \mathbf{y}^{(j-2)}
    \right)
\end{equation}

Moreover, as discussed in \cite{duan2018limited}, to ensure the continuity of the optimized input, an overlapping window, denoted by the superscript $[j]$, is required; it is defined as 
\begin{equation} \label{def_window}
    \mathbf{y}^{[j]} \triangleq \begin{bmatrix} y(jN),\; y(jN+1),\; \ldots,\; y(jN+N_w-1) \end{bmatrix}^\top
\end{equation}

\noindent where $N_w$ is the window length; without loss of generality, $N_w$ in this paper is defined to be twice of the batch length $N$, i.e., $N_w=2N$. Therefore, $\mathbf{\hat{y}}_{h}^{(j+1)}$ also needs to be predicted for control purposes; it can be expressed by
\begin{equation} \label{lhm_preview}
    \mathbf{\hat{y}}_{h}^{(j+1)} =  
    \mathbf{G}_{dd,3}   
    \left(
    \begin{bmatrix} \mathbf{\hat{y}}_{pb}^{(j-2)} \\ \mathbf{\hat{y}}_{pb}^{(j-1)} \\ \mathbf{\hat{y}}_{pb}^{(j)} \end{bmatrix}, 
    \mathbf{\hat{y}}_{pb}^{(j+1)}, \mathbf{y}^{(j-2)}
    \right)
\end{equation}

Combining Eqs.~\eqref{lhm_delay} and \eqref{lhm_preview}, one can write the hybrid model prediction in the $j$-th window, denoted by $\mathbf{\hat{y}}_h^{[j]}$, as
\begin{equation} \label{lhm_window}
    \mathbf{\hat{y}}_h^{[j]} = 
    \begin{bmatrix} \mathbf{\hat{y}}_{h}^{(j)} \\ \mathbf{\hat{y}}_{h}^{(j+1)} \end{bmatrix} =
    \mathbf{G}_{dd}^{[j]} 
    \left(
    \begin{bmatrix} \mathbf{\hat{y}}_{pb}^{(j-2)} \\ \mathbf{\hat{y}}_{pb}^{(j-1)} \end{bmatrix}, 
    \begin{bmatrix} \mathbf{\hat{y}}_{pb}^{(j)} \\ \mathbf{\hat{y}}_{pb}^{(j+1)} \end{bmatrix},
    \mathbf{y}^{(j-2)}
    \right)
\end{equation}

\noindent where $\mathbf{G}_{dd}^{[j]}$ denotes the trained data-driven model used to compute $\mathbf{\hat{y}}_h^{[j]}$. Eq.~\eqref{lhm_window} can be further expressed as a linear operation, similar to Eq.~\eqref{lhm_matrixform}, as
\begin{equation} \label{lhm_win_matrixform}
    \mathbf{\hat{y}}_h^{[j]} =
    \mathbf{L}_{dd,a}^{[j]} \mathbf{\hat{y}}_{pb}^{[j]} +
    \mathbf{L}_{dd,uy}^{[j]} \mathbf{\hat{y}}_{pb}^{[j-1]} +
    \mathbf{L}_{dd,ue}^{[j]} \mathbf{e}_{pb}^{[j-1]} +
    \mathbf{L}_{dd,u1}^{[j]}
\end{equation}

\noindent where $\mathbf{\hat{y}}_{pb}^{[j]}$, $\mathbf{\hat{y}}_{pb}^{[j-1]}$, and $\mathbf{e}_{pb}^{[j-1]}$ respectively represent
\begin{equation} \label{window_to_batch}
    \mathbf{\hat{y}}_{pb}^{[j]} = \begin{bmatrix}  \mathbf{\hat{y}}_{pb}^{(j)} \\ \mathbf{\hat{y}}_{pb}^{(j+1)} \end{bmatrix}, \;
    \mathbf{\hat{y}}_{pb}^{[j-1]} = \begin{bmatrix}  \mathbf{\hat{y}}_{pb}^{(j-2)} \\ \mathbf{\hat{y}}_{pb}^{(j-1)} \end{bmatrix}, \;
    \mathbf{e}_{pb}^{[j-1]} = \mathbf{e}_{pb}^{(j-2)}
\end{equation}

\noindent while $\mathbf{L}_{dd,a}^{[j]}$, $\mathbf{L}_{dd,uy}^{[j]}$, $\mathbf{L}_{dd,ue}^{[j]}$, and $\mathbf{L}_{dd,u1}^{[j]}$ here are constructed based on $\mathbf{\hat{w}}^{(j-1)}$ (due to the unavailability of $\mathbf{y}^{(j-1)}$) and can be computed by replacing $\mathbf{e}_{pb}^{(j-1)}$ with $(\mathbf{\hat{y}}_h^{(j-1)} - \mathbf{\hat{y}}_{pb}^{(j-1)})$ and then comparing the coefficients of the linear operators.

To apply the prediction of Eq.~\eqref{lhm_win_matrixform} to the FBF approach, $\mathbf{\hat{y}}_{h}^{[j]}$ needs to be expressed as a function of the control inputs. The FBF approach formulates the control inputs as the linear combination of a set of basis functions with unknown coefficients, denoted as $\bm{\upgamma}$. Accordingly, the term in Eq.~\eqref{lhm_win_matrixform} that is affected by the optimized inputs, i.e., $\mathbf{\hat{y}}_{pb}^{[j]}$, is represented as a function of $\bm{\upgamma}$. Furthermore, when implemented in receding horizon, as in \cite{duan2018limited}, $\mathbf{\hat{y}}_{pb}$ can be decomposed into past, current, and future portions with respected to window $j$, i.e.,
\begin{equation} \label{def_PCF}
    \begin{aligned}
    \mathbf{\hat{y}}_{pb,P}^{[j]} &\triangleq \begin{bmatrix} y_{pb}(0),\; y_{pb}(1),\; \ldots,\; y_{pb}(jN-1) \end{bmatrix}^\top\\
    \mathbf{\hat{y}}_{pb,C}^{[j]} &\triangleq \mathbf{\hat{y}}_{pb}^{[j]} \\
    \mathbf{\hat{y}}_{pb,F}^{[j]} &\triangleq \begin{bmatrix} y_{pb}(jN+N_w),\; y_{pb}(jN+N_w+1),\; \ldots, \end{bmatrix}^\top
    \end{aligned}
\end{equation}

\noindent where $P$, $C$, and $F$ respectively represent the past, current and future. Accordingly, $\mathbf{\hat{y}}_{pb}$ is formulated as
\begin{equation} \label{FBS_matrix}
    \begin{bmatrix}
        \mathbf{\hat{y}}_{pb,P}^{[j]} \\ \mathbf{\hat{y}}_{pb,C}^{[j]} \\ \mathbf{\hat{y}}_{pb,F}^{[j]}
    \end{bmatrix} =
    \begin{bmatrix}
        \mathbf{\tilde{\Psi}}_{P} & \mathbf{0} & \mathbf{0} \\
        \mathbf{\tilde{\Psi}}_{PC} & \mathbf{\tilde{\Psi}}_{C} & \mathbf{0} \\
        \mathbf{\tilde{\Psi}}_{PF} & \mathbf{\tilde{\Psi}}_{CF} & \mathbf{\tilde{\Psi}}_{F}
    \end{bmatrix}
    \begin{bmatrix}
        \bm{\upgamma}_P^{[j]} \\ \bm{\upgamma}_C^{[j]} \\ \bm{\upgamma}_F^{[j]}
    \end{bmatrix}
\end{equation}

\noindent where $\bm{\upgamma}_P^{[j]}$, $\bm{\upgamma}_C^{[j]}$, $\bm{\upgamma}_F^{[j]}$ represent the vectors of coefficients corresponding to past, current, and future windows, $\mathbf{\tilde{\Psi}}$ matrices represent the concatenation of the basis functions that are filtered by  $\mathbf{G}_{pb}$, with the subscripts $P$, $C$, $F$ of $\mathbf{\tilde{\Psi}}$ denoting the input-output effects in the past, current and future sections, and the subscripts $PC$ denoting the effects of the past inputs on the current outputs, similarly for $PF$ and $CF$. Also, note that $\mathbf{\tilde{\Psi}}$ is time-invariant when the batch length $N$ is a multiple of the number of time steps represented by one coefficient ${\upgamma}$.   

From Eqs.~\eqref{def_PCF} and \eqref{FBS_matrix}, $\mathbf{\hat{y}}_{pb}^{[j]}$ is represented as a function of $\bm{\upgamma}$ by
\begin{equation} \label{ypb_curr}
    \mathbf{\hat{y}}_{pb}^{[j]} = 
    \mathbf{\tilde{\Psi}}_{C}  \bm{\upgamma}_C^{[j]} +
    \mathbf{\tilde{\Psi}}_{PC} \bm{\upgamma}_P^{[j]}
\end{equation}
which is then inserted to Eq.~\eqref{lhm_win_matrixform}. Replacing the unalterable terms in Eq.~\eqref{lhm_win_matrixform}, i.e., the terms from the past ($\mathbf{\hat{y}}_{pb}^{[j-1]}$ and $\mathbf{e}_{pb}^{[j-1]}$) and the bias term with $\mathbf{\Phi}_u^{[j]}$, and concatenating their linear operators as $\mathbf{L}_{dd,u}^{[j]}$, one can formulate $\mathbf{\hat{y}}_h^{[j]}$ as
\begin{equation} \label{lhm_final}
    \mathbf{\hat{y}}_h^{[j]} =
    \mathbf{L}_{dd,a}^{[j]} \mathbf{\tilde{\Psi}}_{C}  \bm{\upgamma}_C^{[j]} +
    \mathbf{L}_{dd,a}^{[j]} \mathbf{\tilde{\Psi}}_{PC} \bm{\upgamma}_P^{[j]} +
    \mathbf{L}_{dd,u}^{[j]} \mathbf{\Phi}_u^{[j]}
\end{equation}

Finally, the tracking error is minimized window by window \cite{duan2018limited}. Given a desired trajectory for the $j$-th window $\mathbf{y}_d^{[j]}$, since $\bm{\upgamma}_C^{[j]}$ is the only optimization variable, the objective function is given by
\begin{equation} \label{HybridFBF_opt}
    \underset {\bm{\upgamma}_C^{[j]}} {\text{min}}
        \left\| \mathbf{e}^{[j]} \right\|^2_2
    \underset {\bm{\upgamma}_C^{[j]}} {\text{min}}
        \left\| \mathbf{y}_d^{[j]} - \mathbf{\hat{y}}_h^{[j]} \right\|^2_2
\end{equation}

\noindent Therefore, the optimal $\bm{\upgamma}_C^{[j]}$ for the hybrid FBF controller is calculated as
\begin{equation} \label{HybridFBF_main}
    \bm{\upgamma}_C^{[j]} = 
    \left( \mathbf{L}_{dd,a}^{[j]} \mathbf{\tilde{\Psi}}_{C} \right)^{\dagger} 
    \left( \mathbf{y}_d^{[j]} 
    - \mathbf{L}_{dd,a}^{[j]} \mathbf{\tilde{\Psi}}_{PC} \bm{\upgamma}_P^{[j]} 
    - \mathbf{L}_{dd,u}^{[j]} \mathbf{\Phi}_u^{[j]}
    \right)
\end{equation}

\noindent Then, as in \cite{duan2018limited}, the optimized control input for the $j$-th batch $\mathbf{u}^{(j)}$ is reconstructed by the linear combination of the basis functions, with only the first $N$ time steps extracted, i.e.,
\begin{equation} \label{u_curr}
    \mathbf{u}^{(j)} = 
    \begin{bmatrix} \mathbf{I}_N & \mathbf{0} \end{bmatrix}
    \left( \mathbf{\Psi}_{C}  \bm{\upgamma}_C^{[j]} + \mathbf{\Psi}_{PC} \bm{\upgamma}_P^{[j]} \right)
\end{equation}

After the optimized $\mathbf{u}^{(j)}$ is fed to the system, and the corresponding output $\mathbf{y}^{(j)}$ is measured, the linear hybrid model $\mathbf{G}_h$ is further updated by Eq.~\eqref{lhm_train} using recursive least squares algorithm, with additional $e_{pb}(k)$ terms coming from
\begin{equation}
    \begin{aligned}
    \mathbf{e}_{pb}^{(j)} 
    &= \mathbf{y}^{(j)} - \mathbf{\hat{y}}_{pb}^{(j)} \\
    &= \mathbf{y}^{(j)} - \begin{bmatrix} \mathbf{I}_N & \mathbf{0} \end{bmatrix}
        \left( \mathbf{\tilde{\Psi}}_{C}  \bm{\upgamma}_C^{[j]} + \mathbf{\tilde{\Psi}}_{PC} \bm{\upgamma}_P^{[j]} \right)
    \end{aligned}
\end{equation}

\noindent A flowchart for the hybrid FBF controller is shown in Fig.~\ref{hybridfbf}.

\noindent {\bf{Remark 3:}} When the model weight $\mathbf{\hat{w}}$ of the linear hybrid model $\mathbf{G}_h$ is set equal to zero, $\mathbf{L}_{dd,a}^{[j]}$ and $\mathbf{L}_{dd,u}^{[j]}$ in Eq.~\eqref{lhm_final} will respectively become an identity matrix and a zero matrix, thereby reducing the hybrid FBF approach to the standard FBF method of \cite{duan2018limited}. Under this scenario, it can also be inferred from Eq.~\eqref{lhm_recursive} that $\mathbf{\hat{y}}_{h}=\mathbf{\hat{y}}_{pb}$.

\section{Stability Analysis for Hybrid FBF Controller} \label{stability}

Although the linear hybrid model can give more accurate prediction \cite{chou2021linear} and accordingly lead to more accurate tracking using the hybrid FBF controller, the hybrid FBF controller may potentially suffer from instability due to the feedback loop introduced by the regression on past data. More specifically, according to Eq.~\eqref{HybridFBF_main}, the feedback of the $\mathbf{\Phi}_u^{[j]}$ term, which includes the past outputs, may potentially cause the overall closed-loop dynamics to become unstable. Furthermore, the varying values of the $\mathbf{L}_{dd}$ matrices in Eq.~\eqref{HybridFBF_main} due to training can also affect the stability of the system.

Therefore, to ensure the safety of the controller, we propose an analytical approach to check the stability of the system when using the hybrid FBF controller in this section. The stability of the hybrid FBF approach largely depends on the measured data from the past. Therefore, the stability analysis proposed in this section aims to check if there will be any risk of instability for the next batch, given the trained results of $\mathbf{\hat{w}}$. In practice, once impending instability is detected, a variety of mitigating actions can be taken to prevent it. For example, learning could be turned off, allowing the system to revert to the standard FBF controller, or the system could be run in a safe mode using conservative speed and acceleration.

To perform the stability check, we assume that the linear hybrid model has converged to accurate predictions, and thus the rate of weight changes due to updating is sufficiently low. This assumption is reasonable because, in practice, the hybrid model should be adequately trained for a few cycles to ensure that it is sufficiently accurate before using it for control. Given this assumption, the general idea of the stability check method is to write the linear hybrid model in the state-space format as the approximation of the actual system, i.e., approximating $\mathbf{y}=\mathbf{\hat{y}}_h$ , and then formulating the input to the system as a control feedback law utilizing the hybrid FBF controller. That is, the linear hybrid model, described in Eq.~\eqref{lhm_win_matrixform}, and the hybrid FBF controller, formulated as Eq.~\eqref{HybridFBF_main}, are rewritten into the form of
\begin{equation} \label{state_space}
    \begin{split}
    \mathbf{x}^{[j+1]} &= \mathbf{A}^{[j]} \mathbf{x}^{[j]} + \mathbf{B}^{[j]} \mathbf{u}^{[j]} \\
    \mathbf{u}^{[j]}   &= \mathbf{K}^{[j]} \mathbf{x}^{[j]} + \mathbf{M}^{[j]} \mathbf{r}^{[j]}
    \end{split}
\end{equation} 

\noindent where the input vector $\mathbf{u}^{[j]}$ and the reference vector $\mathbf{r}^{[j]}$ are represented by the FBF coefficients $\bm{\upgamma}_C^{[j]}$ and the desired trajectory $\mathbf{y}_d^{[j]}$. For the design of the state vector, $\mathbf{x}^{[j]}$, based on Eq.~\eqref{lhm_win_matrixform} and Eq.~\eqref{ypb_curr}, $\mathbf{\hat{y}}_{pb}$ and $\mathbf{\hat{y}}_h$ from \mbox{$(j-2)$-th} to \mbox{$j$-th} batch, as well as $\bm{\upgamma}_P^{[j]}$ and the bias term must be included. Accordingly, $\mathbf{x}^{[j]}$, $\mathbf{u}^{[j]}$, and $\mathbf{r}^{[j]}$ are designed as
\begin{subequations} \label{ss_vectors}
\begin{align}
    &\begin{aligned}
        \mathbf{x}^{[j]} = \Big[  
        & {{}\mathbf{\hat{y}}_h^{(j-2)}}^\top , \; {{}\mathbf{\hat{y}}_h^{(j-1)}}^\top, \; {{}\mathbf{\hat{y}}_h^{(j)}}^\top,\\
        & \; {{}\mathbf{\hat{y}}_{pb}^{(j-2)}}^\top, \; {{}\mathbf{\hat{y}}_{pb}^{(j-1)}}^\top, \; {{}\mathbf{\hat{y}}_{pb}^{(j)}}^\top, \; {\bm{\upgamma}^{[j]}}^\top, \; 1
        \Big]^\top
    \end{aligned} \\
    &\mathbf{u}^{[j]} = {\bm{\upgamma}^{[j]}}^\top = \begin{bmatrix} {\bm{\upgamma}^{(j)}}^\top, & {\bm{\upgamma}^{(j+1)}}^\top \end{bmatrix}^\top \\
    &\mathbf{r}^{[j]} = \mathbf{y}_d^{[j]} = \begin{bmatrix} {{}\mathbf{y}_d^{(j)}}^\top, & {{}\mathbf{y}_d^{(j+1)}}^\top \end{bmatrix}^\top
\end{align}
\end{subequations}

\noindent Furthermore, by comparison of the linear operators of Eq.~\eqref{state_space} with that of Eqs.~\eqref{lhm_win_matrixform}, \eqref{ypb_curr}, \eqref{lhm_final}, and \eqref{HybridFBF_main}, the matrices $\mathbf{A}^{[j]}$, $\mathbf{B}^{[j]}$, $\mathbf{K}^{[j]}$, and $\mathbf{M}^{[j]}$ are given by
\begin{subequations} \label{ss_matrices}
\begin{align}
    \mathbf{A}^{[j]} &= 
    \begin{bmatrix} 
    \mathbf{0}      & \mathbf{I}_N & \mathbf{0} & \mathbf{0}      & \mathbf{0}      & \mathbf{0} & \mathbf{0}      & \mathbf{0} \\
    \mathbf{A}_{21} & \mathbf{0}   & \mathbf{0} & \mathbf{A}_{24} & \mathbf{A}_{25} & \mathbf{0} & \mathbf{A}_{27} & \mathbf{A}_{28} \\
    \mathbf{A}_{31} & \mathbf{0}   & \mathbf{0} & \mathbf{A}_{34} & \mathbf{A}_{35} & \mathbf{0} & \mathbf{A}_{37} & \mathbf{A}_{38} \\
    \mathbf{0}      & \mathbf{0}   & \mathbf{0} & \mathbf{0}      & \mathbf{I}_N    & \mathbf{0} & \mathbf{0}      & \mathbf{0} \\
    \mathbf{0}      & \mathbf{0}   & \mathbf{0} & \mathbf{0}      & \mathbf{0}      & \mathbf{0} & \mathbf{A}_{57} & \mathbf{0} \\
    \mathbf{0}      & \mathbf{0}   & \mathbf{0} & \mathbf{0}      & \mathbf{0}      & \mathbf{0} & \mathbf{A}_{67} & \mathbf{0} \\
    \mathbf{0}      & \mathbf{0}   & \mathbf{0} & \mathbf{0}      & \mathbf{0}      & \mathbf{0} & \mathbf{A}_{77} & \mathbf{0} \\
    \mathbf{0}      & \mathbf{0}   & \mathbf{0} & \mathbf{0}      & \mathbf{0}      & \mathbf{0} & \mathbf{0}      & \mathbf{0}
    \end{bmatrix}  \\
    \mathbf{B}^{[j]} &= 
    \begin{bmatrix}
    \mathbf{0} & \mathbf{B}_{2}^\top & \mathbf{B}_{3}^\top & \mathbf{0} & \mathbf{B}_{5}^\top & \mathbf{B}_{6}^\top & \mathbf{B}_{7}^\top & \mathbf{0}
    \end{bmatrix}^\top  \\
    \mathbf{K}^{[j]} &= 
    \begin{bmatrix}
    \mathbf{K}_{1} & \mathbf{0} & \mathbf{0} & \mathbf{K}_{4} & \mathbf{K}_{5} & \mathbf{0} & \mathbf{K}_{7} & \mathbf{K}_{8}
    \end{bmatrix} \\
    \mathbf{M}^{[j]} &= (\mathbf{L}_{dd,a}^{[j]} \mathbf{\tilde{\Psi}}_{C})^{\dagger}
\end{align}
\end{subequations}

\noindent where $\mathbf{I}_N$ are N-by-N identity matrices, $\mathbf{0}$ are zero matrices with appropriate size, and the submatrices $\mathbf{A}_{ij}$, $\mathbf{B}_{ij}$, and $\mathbf{K}_{ij}$ are defined as follows:
\begin{align}
    & \begin{bmatrix} \mathbf{A}_{21} \\ \mathbf{A}_{31} \end{bmatrix} = \mathbf{L}_{dd,ue}^{[j]}, \quad 
    \begin{bmatrix} \mathbf{A}_{28} \\ \mathbf{A}_{38} \end{bmatrix} = \mathbf{L}_{dd,u1}^{[j]},  \nonumber \\
    & \begin{bmatrix} \mathbf{A}_{24} \\ \mathbf{A}_{34} \end{bmatrix} = \mathbf{L}_{dd,uy}^{[j]} \begin{bmatrix} \mathbf{I}_{N} \\ \mathbf{0} \end{bmatrix} -\mathbf{L}_{dd,ue}^{[j]}, \nonumber \\
    & \begin{bmatrix} \mathbf{A}_{25} \\ \mathbf{A}_{35} \end{bmatrix} = \mathbf{L}_{dd,uy}^{[j]} \begin{bmatrix} \mathbf{0} \\ \mathbf{I}_{N} \end{bmatrix}, \nonumber \\
    & \begin{bmatrix} \mathbf{A}_{27} \\ \mathbf{A}_{37} \end{bmatrix} = \mathbf{L}_{dd,a}^{[j]} \mathbf{\tilde{\Psi}}_{PC}, \quad 
    \begin{bmatrix} \mathbf{B}_{2} \\ \mathbf{B}_{3} \end{bmatrix}   = \mathbf{L}_{dd,a}^{[j]} \mathbf{\tilde{\Psi}}_{C}, \nonumber \\
    &\begin{bmatrix} \mathbf{A}_{57} \\ \mathbf{A}_{67} \end{bmatrix} = \mathbf{\tilde{\Psi}}_{PC}, \quad 
    \begin{bmatrix} \mathbf{B}_{5} \\ \mathbf{B}_{6} \end{bmatrix}   = \mathbf{\tilde{\Psi}}_{C}, \nonumber \\
    &\mathbf{A}_{77} = \begin{bmatrix} \mathbf{0} & \mathbf{I}_{n_p-n_c} \\ \mathbf{0} & \mathbf{0} \end{bmatrix}, \quad 
    \mathbf{B}_{7}  = \begin{bmatrix} \mathbf{0} & \mathbf{0} \\ \mathbf{I}_{n_c} & \mathbf{0} \end{bmatrix}, \nonumber \\
    & \mathbf{K}_{i} = -(\mathbf{L}_{dd,a}^{[j]} \mathbf{\tilde{\Psi}}_{C})^{\dagger} \begin{bmatrix} \mathbf{A}_{2i} \\ \mathbf{A}_{3i} \end{bmatrix} \text{ (for } i=1,4,5,7,8 \text{)}
\end{align}
where $n_p$ and $n_c$ are the sizes of $\bm{\upgamma}_P^{[j]}$ and $\bm{\upgamma}_C^{[j]}$, respectively.

Finally, based on the assumption that the linear hybrid model is converging, i.e., the overall dynamics described in Eq.~\eqref{state_space} can be assumed to be a slow linear time-varying system before instability happens due to the low rate of weight changes (validated in Section~\ref{exp_stability}), the closed-loop dynamics can be approximated as a linear time-invariant (LTI) system. Accordingly, based on LTI stability analysis, it can be determined that the hybrid FBF controller is unstable when the closed-loop state matrix $(\mathbf{A}^{[j]} + \mathbf{B}^{[j]} \mathbf{K}^{[j]})$ has eigenvalues outside the unit disk. Hence, instability is predicted to occur in the next batches when the spectral radius (i.e., largest absolute eigenvalue) exceeds 1.

\section{Application to 3D Printing} \label{experiments}
In this section, the proposed hybrid FBF controller and stability analysis approach are validated experimentally on a desktop 3D printer. 

\subsection{Background}
\label{app_background}
Desktop 3D printers are typically made of lightweight material and are driven by stepper motors through belts. As a result, they experience unwanted vibration during printing, leading to a loss of printed part quality. A standard FBF controller, with no data-driven component, has been successfully used to compensate the unwanted vibration, enabling high-speed printing while maintaining high print quality \cite{duan2018limited, edoimioya2021software}. However, the physics-based model used in the standard FBF controller does not capture the nonlinear vibration behavior of desktop 3D printers, e.g., due to nonlinear friction, backlash, and belt stiffness. Moreover, 3D printers wear over time, and it is not unusual for end-users to perform aftermarket modifications to their desktop printers, e.g., adding a new nozzle or installing a webcam, thus altering the printers' dynamics in unknown ways. The standard FBF approach is unable to handle these unmodeled dynamics. However, next-generation desktop 3D printers are being equipped with low-cost accelerometers that can measure vibration. In the following subsections, we explore the effectiveness of the hybrid FBF controller to leverage data gathered from on-board accelerometers to compensate unmodeled vibration dynamics of a desktop 3D printer.

\subsection{Experimental Setup} \label{exp_setup}

\begin{figure}[!t]
\centering
\includegraphics[width=\columnwidth]{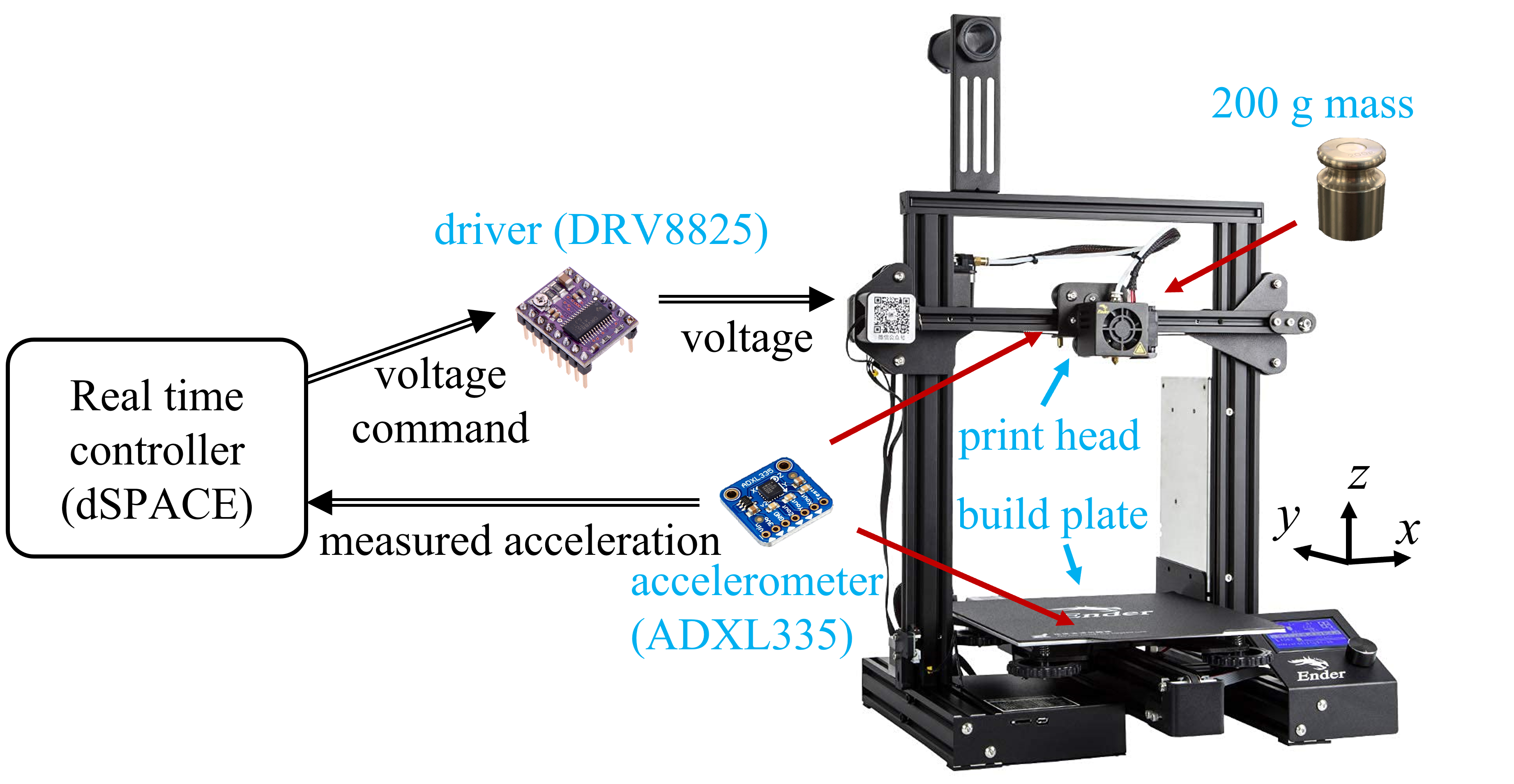}
\caption{The printer and controller setup used for the experimental case studies (the 200 g mass is added for case study 2 and 3 only).}
\label{Printer_setup}
\end{figure}

\begin{figure}[!t]
\centering
\includegraphics[width=\columnwidth]{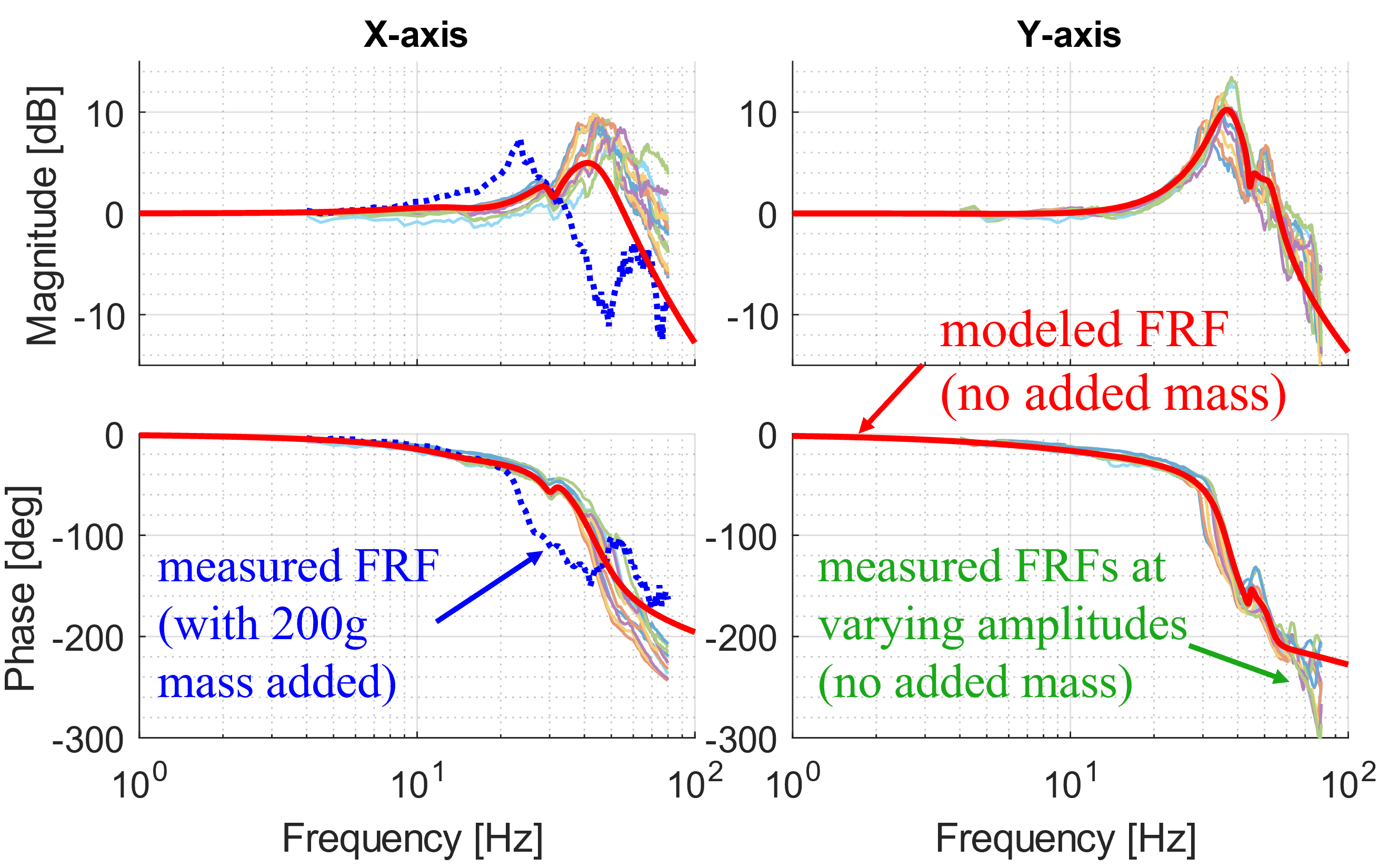}
\caption{Comparison of the FRF of the identified model, the measured FRFs from the unmodified printer (no added mass) using different amplitudes of sine sweeps, and (for $x$-axis only) the measured FRF from the retrofitted printer (with 200 g mass added)}
\label{Exp_frf}
\end{figure}

Figure~\ref{Printer_setup} shows the experimental set up. It consists of a Creality Ender 3 Pro desktop 3D printer. A 200 g mass can be attached to the printer's $x$-axis to simulate an aftermarket modification to the printer (e.g., the installation of a heavier printhead or a webcam). Two low-cost accelerometers (Adafruit ADXL335) are attached to the printer’s $x$- and $y$-axis to collect acceleration signals, which are converted to position signals using a Luenberger observer with a bandwidth of 5 Hz. All control algorithms tested on the printer are implemented on dSPACE DS1107 and dSPACE DS5203 FPGA board at fixed sampling rates of 1 kHz (for control) and 40kHz (for data collection and stepper motor command generation). The stepper motor commands are delivered to the printer's motors using DRV8825 stepper motor drivers.

Figure~\ref{Exp_frf} shows the printer's $x$- and $y$-axis frequency response functions (FRFs). The plot shows FRFs measured at varying input acceleration amplitudes without the 200 g mass added to the printer. Due to the nonlinear dynamics of the printer, the FRFs vary significantly with the input amplitude. Also shown is a linear transfer function model fit using the measured FRFs. The identified transfer functions for the $x$- and $y$-axes are given by
\begin{equation}
    \mathbf{G}_{pb,x} = \frac{\text{num}_x(s)}{\text{den}_x(s)}
    \quad \text{and} \quad
    \mathbf{G}_{pb,y} = \frac{\text{num}_y(s)}{\text{den}_y(s)}
\end{equation}
where
\begin{align*}
    &\text{num}_x(s) = -62.48 s^5 + 5.91\times10^4 s^4 + 3.82\times10^6 s^3 + \\*
        & 2.96\times10^9 s^2 + 1.96\times10^{11} s + 2.29\times10^{13} \\
    &\text{den}_x(s) = s^6 + 242.6 s^5 + 1.36\times10^5 s^4 + 1.73\times10^7 s^3 +\\*
        & 4.22\times10^9 s^2 + 2.75\times10^{11} s + 2.29\times10^{13} \\
    &\text{num}_y(s) = -84.79 s^6 + 2.87\times10^4 s^5 - 8.03\times10^6 s^4 +\\*
        & 6.45\times10^9 s^3 + 3.86\times10^{11} s^2 + 3.29\times10^{14} s + 3.74\times10^{16} \\
    &\text{den}_y(s) = s^7 + 211.2 s^6 + 2.56\times10^5 s^5 + 4.11\times10^7 s^4  +\\*
        & 2.07\times10^{10} s^3 + 2.39\times10^{12} s^2 + 5.28\times10^{14} s + 3.74\times10^{16}
\end{align*}

Notice that the fit FRF, which is used to generate the physics-based model $\mathbf{G}_{pb}$ used in the standard and hybrid FBF controllers, does not model the nonlinear amplitude-dependent behavior of the FRFs. Instead, it models the average of all measured FRFs, with larger weighting on the lower-frequency signals, which is more common for most trajectories. As a result, in addition to the amplitude-dependent behavior, the fit FRF and the measured FRFs may also have larger deviation at higher frequencies. Also shown on the $x$-axis plot is an FRF measured when the 200 g mass is added to $x$-axis. Notice that this modification causes the printer's FRF to further deviate from the modeled FRF. Two case studies are described below that evaluate the effectiveness of the hybrid FBF controller in handling these unmodeled dynamics. A third case study evaluates the stability of the controller. The control parameters used in the three case studies are summarized in Table~\ref{hybridfbf_params}.

\begin{table}
\begin{center}
\caption{The controller parameters of the standard and the hybrid FBF used for the experimental case studies. }
\label{hybridfbf_params}
\begin{tabular}{| l | c |}
\hline
\multirow{2}{0.4\linewidth}{Sampling rate} & {1 kHz (for control)} \\
    & {40 kHz (for data transmission)} \\
\hline
{Observer bandwidth} & {5 Hz} \\
\hline
{Accelerometers low-pass cutoff freq.} & {200 Hz} \\
\hline
{B-spline degree} & {5} \\
\hline
{B-spline knot vector spacing} & {10 time steps (= 0.01 s)} \\
\hline
{$N$ (batch length)} & {70 time steps (= 0.07 s)} \\
\hline
{$N_w$ (window length)} & {140 time steps (= $2N$)} \\
\hline
{$q$ (number of $y_{pb}$ terms in $\bm{\upphi}$)} & {4} \\ 
\hline
{$p$ (number of $e_{pb}$ terms in $\bm{\upphi}$)} & {50} \\
\hline
{$\lambda$ (regularization factor)} & {0.01} \\
\hline
\end{tabular}
\end{center}
\end{table}

\subsection{Case Study 1: Evaluation of Hybrid FBF Controller on Unmodified Printer} \label{exp_case1}

\begin{figure}[!t]
\centering
\includegraphics[width=\columnwidth]{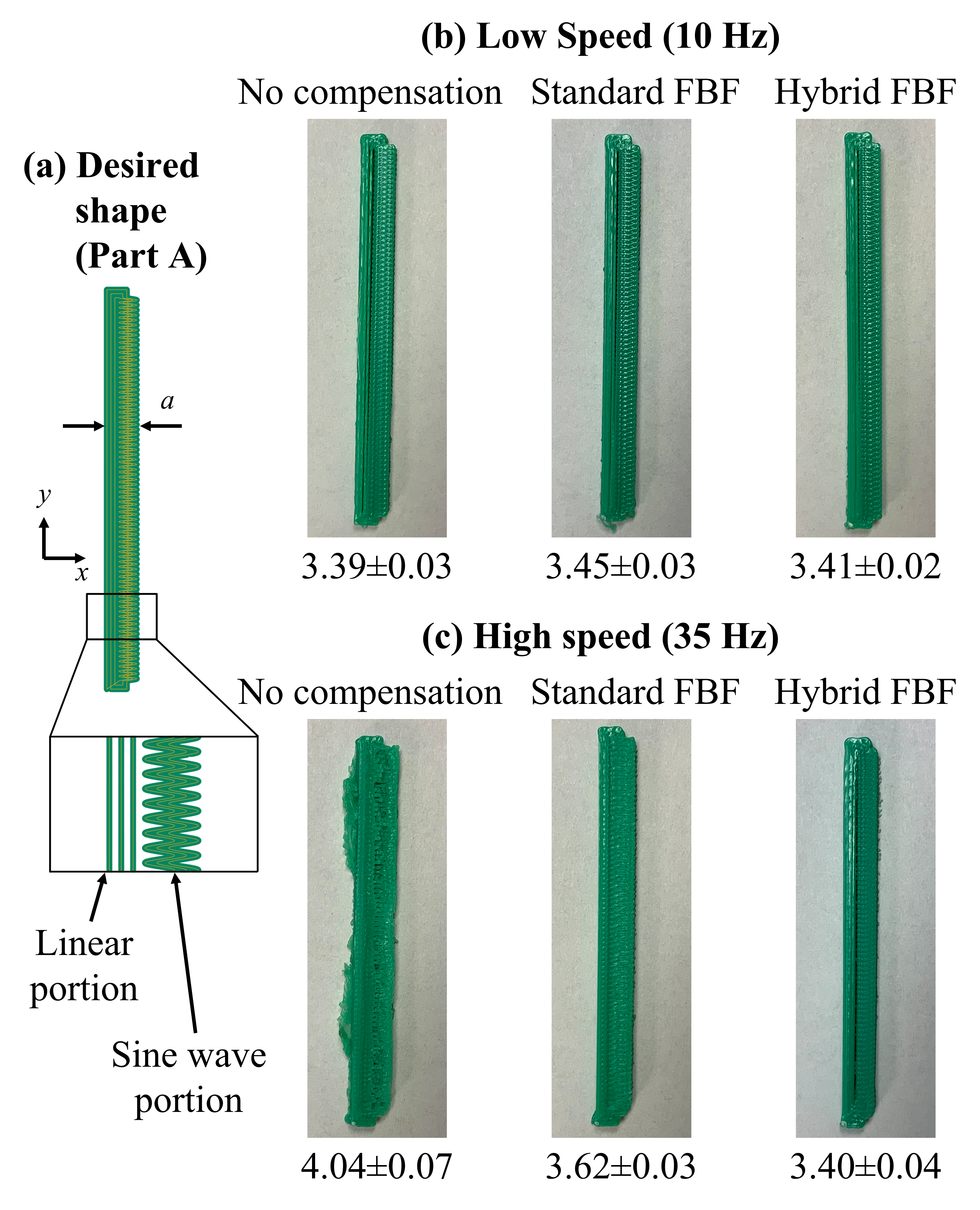}
\caption{The (a) desired shape for case study 1 (i.e., Part A) and the comparison of the quality of parts printed at (b) low speed and (c) high speed using different controllers. (The numbers show the mean width $a$ and its standard deviation of the parts from 10 measured values, in mm.)}
\label{Exp1_prints}
\end{figure}

\begin{figure}[!t]
\centering
\includegraphics[width=\columnwidth]{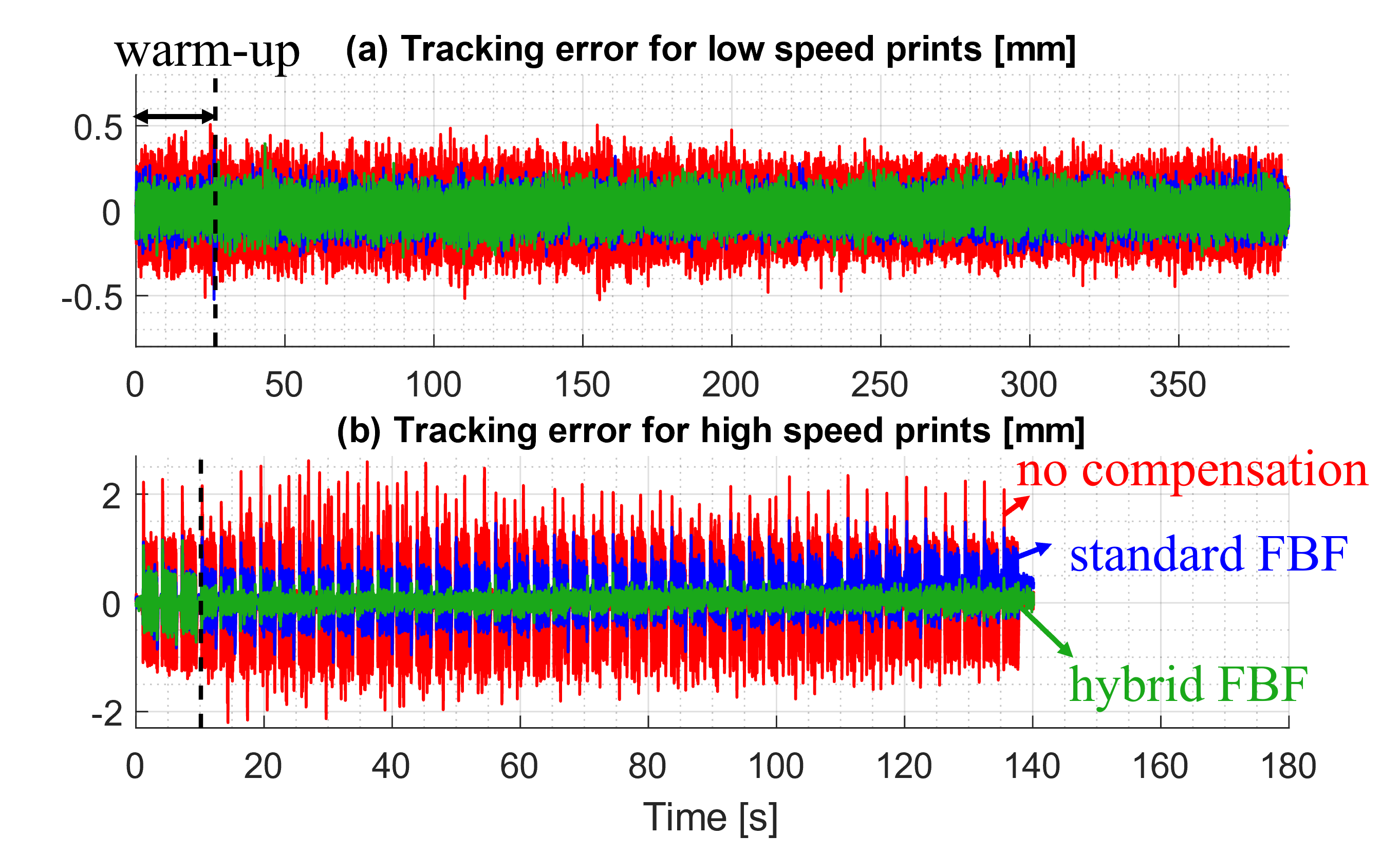}
\caption{Comparison of the $x$-axis tracking error using different controllers for (a) low-speed and (b) high-speed printing for case study 1.}
\label{Exp1_results}
\end{figure}

This case study aims to show the quality improvement of 3D printed parts using the hybrid FBF controller on the printer of Fig.~\ref{Printer_setup} without the 200 g mass added. The unmodeled dynamics are, therefore, due to the nonlinear amplitude-dependent behavior shown in Fig.~\ref{Exp_frf}. Part A, shown in Fig.~\ref{Exp1_prints}(a), is printed without compensation and with compensation using the standard and hybrid FBF controllers. The part consists of linear and sine-wave motions. The sine wave motions have a wavelength of 0.5 mm. Therefore, when traversing the sine wave, the $x$-axis vibrates at a frequency proportional to the feed rate of the printer, and the frequency of the sine wave can be obtained by: frequency = ($y$-axis speed)/wavelength.

Figure~\ref{Exp1_prints}(b) and (c) show the results of Part A with the sine wave printed at low speed (i.e., 10~Hz) and high speed (i.e., 35~Hz) without compensation, and with compensation using the standard and hybrid FBF controllers. The width $a$ shown in Fig.~\ref{Exp1_prints}(a) is used as a figure of merit. It is obtained by measuring  10 different locations of the printed parts using a vernier caliper, and computing the mean and the standard deviation as summarized in Fig.~\ref{Exp1_prints}(b) and (c). At low speed, all three prints have similar quality, as shown by their similar mean values of $a$. This is because, as seen from Fig.~\ref{Exp_frf}, at 10~Hz, the resonance of the machine is not excited, and there is hardly any vibration to compensate. Moreover, the nonlinear behavior of the machine is not prevalent. However, at 35~Hz, the qualities of the parts are markedly different. The mean width of the parts printed without compensation and with compensation using the standard FBF controller are respectively 19.17\% and 4.93\% larger than their low-speed counterparts due to uncompensated vibration. The standard FBF controller does a poor job because it is unable to compensate the unmodeled nonlinear vibration around 35~Hz. However, the hybrid FBF controller does a much better job at compensation, leading to only 0.29\% larger mean value of $a$ at high speed compared to at low speed.

The performance of using the hybrid FBF controller can also be observed from the time history of the tracking error. Here, the estimation of the observer is treated as the true position of the printer. Figure~\ref{Exp1_results} compares the tracking error of the parts printed without compensation and with compensation using the standard and hybrid FBF controllers. At low speed, as shown in Fig.~\ref{Exp1_results}(a), the standard and hybrid FBF controllers have similar levels of tracking error, while that of the uncompensated is larger. At high speed, Fig.~\ref{Exp1_results}(b) shows that the uncompensated and the standard FBF control cases both have significantly larger tracking errors than the hybrid FBF controller beyond the warm-up period, where $\mathbf{G}_{dd}$ is being trained and thereby the standard and hybrid FBF controllers are equivalent. The overall root mean square (RMS) tracking error between the uncompensated, standard FBF, and hybrid FBF are respectively 143.3, 74.7, and 73.2 $\mu$m for low speed printing and 702.6, 358.6, and 129.8 $\mu$m for high-speed printing. Again, this confirms the superior performance of the hybrid FBF controller in compensating unmodeled nonlinear vibration.

\subsection{Case Study 2: Evaluation of Hybrid FBF Controller on Retrofitted Printer} \label{exp_case2}

\begin{figure}[!t]
\centering
\includegraphics[width=\columnwidth]{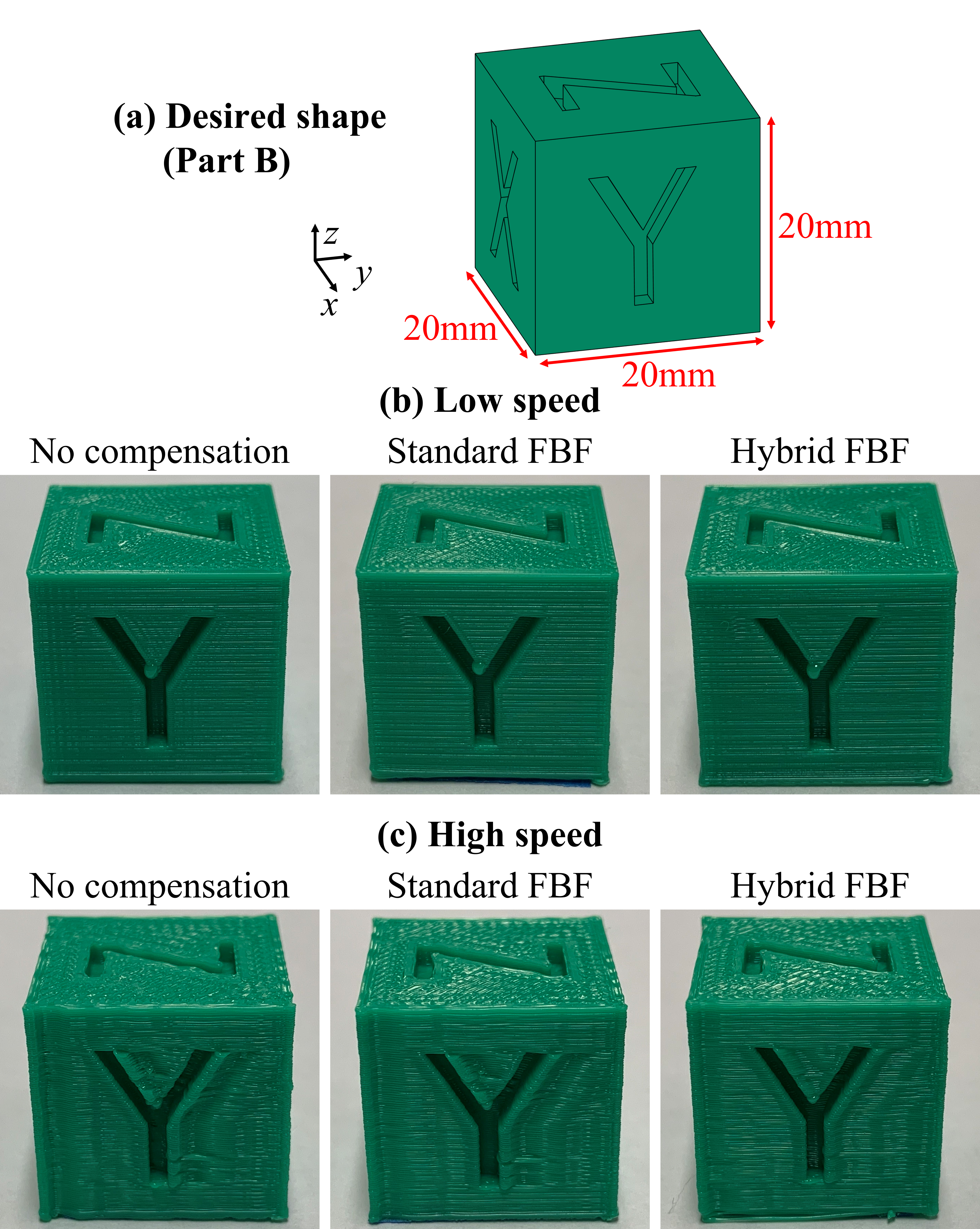}
\caption{The (a) desired shape for case study 2 (i.e., Part B) and the comparison of the print quality using different controllers printed at (b) low speed and (c) high speed}
\label{Exp2_prints}
\end{figure}

\begin{figure}[!t]
\centering
\includegraphics[width=\columnwidth]{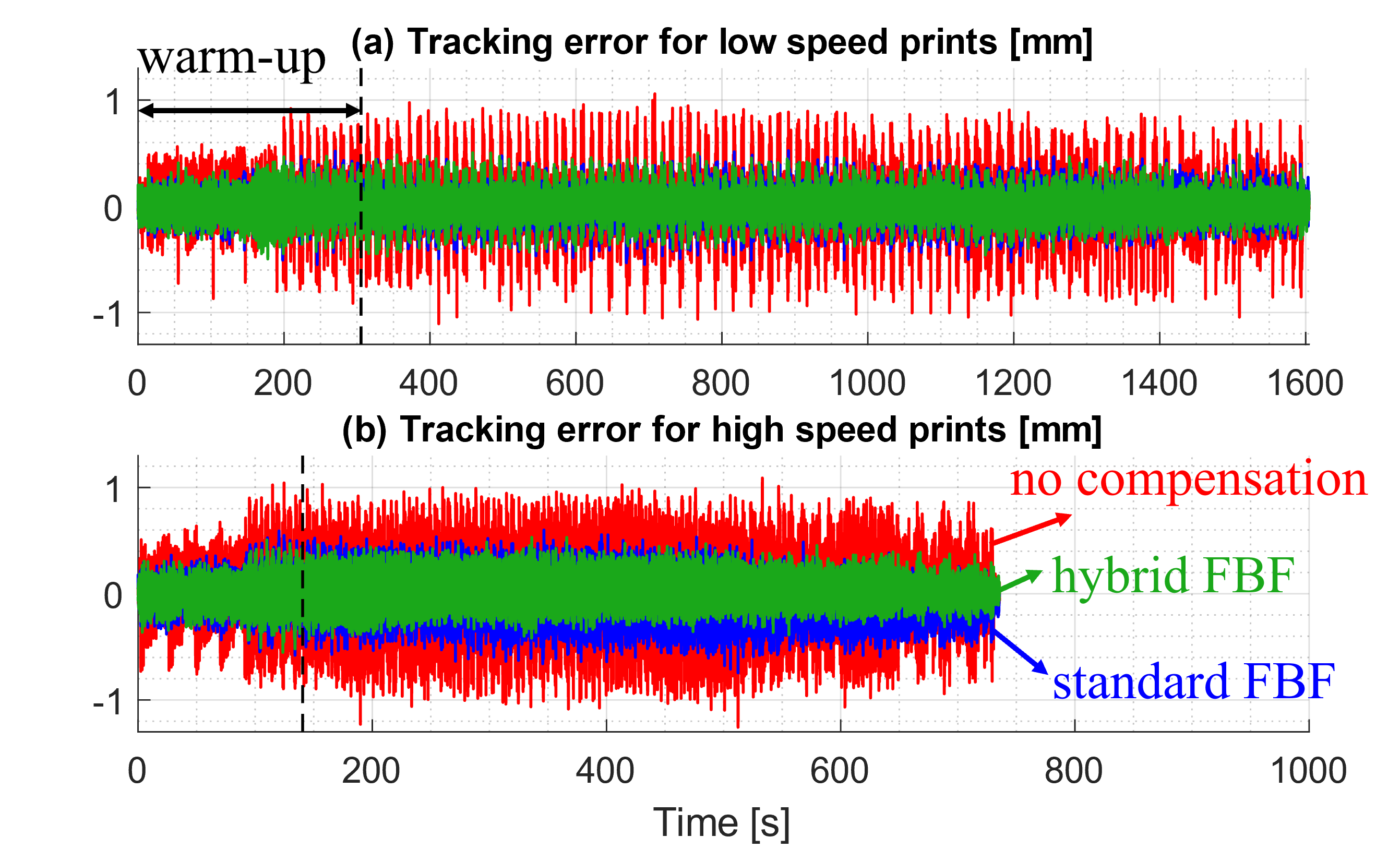}
\caption{Comparison of the $x$-axis tracking error using different controllers for (a) low-speed and (b) high-speed printing for case study 2.}
\label{Exp2_results}
\end{figure}

In this case study, Part B shown in Fig.~\ref{Exp2_prints}(a) is printed on the 3D printer in Fig.~\ref{Printer_setup} with the 200 g mass added. As shown in Fig.~\ref{Exp_frf}, the added mass leads to significant mismatch between the modeled and actual $x$-axis dynamics. Part B is printed without compensation, and with compensation using the standard and hybrid FBF controllers at low speed (i.e., 25 mm/s wall speed) and high speed (i.e., 100 mm/s wall speed). The feed rate, acceleration, and jerk limits for both speed settings are respectively:
\begin{equation}
    v_{\text{lim}} = 100 \,\text{mm}/\text{s}, \; 
    a_{\text{lim}} = 10 \,\text{m}/\text{s}^2, \; 
    j_{\text{lim}} = 5000 \,\text{m}/\text{s}^3
\end{equation}

As shown in Fig.~\ref{Exp2_prints}(b), at low speed, the quality of the part is similar for the three cases. However, at high speed, as shown in Fig.~\ref{Exp2_prints}(c), there is a loss of quality on the “Y” face of the cube due to unmodeled $x$-axis vibration. The loss of quality is severe without compensation. Its severity is slightly reduced using the standard FBF controller. However, the hybrid FBF controller significantly reduces the loss of quality thanks to its ability to leverage data gathered from the on-board accelerometers to learn the unmodeled dynamics. Note that the loss of quality in the high-speed case is not perfectly compensated by the hybrid FBF controller. One reason is that the compensation using the hybrid model is based on position signals observed from acceleration signals using the modeled FRF in Fig.~\ref{Exp_frf}, which is highly inaccurate after retrofitting the printer. This limitation could be alleviated if position signals are gathered directly from a position sensor. Another reason is that the training of the key feature, i.e., the “Y” letter notch, is diluted by other parts of the shape that do not have the same geometric pattern. This problem can be addressed by performing more intelligent training updates of $\mathbf{G}_{dd}$ using features with similar geometric patterns.

Similar to Case Study 1, the quantitative tracking performance of the hybrid FBF is shown in the time history plots of the tracking error in Fig.~\ref{Exp2_results}. As shown in Fig.~\ref{Exp2_results}(a), at low speed, the uncompensated case has some spikes in the tracking error, which are compensated by both the standard and the hybrid FBF. These spikes of tracking error occur mostly in the high-speed infill (interior) printing, and thus are not reflected on the surface quality of the printed part shown in Fig.~\ref{Exp2_prints}(b). However, for high-speed printing, Fig.~\ref{Exp2_results}(b) shows that the uncompensated cases have large tracking error at almost every instant. The standard FBF can slightly compensate it, while the hybrid FBF can further improve the compensation of the standard FBF after the warm-up section. The overall RMS tracking error between the uncompensated, standard FBF, and hybrid FBF are respectively 146.6, 77.5, and 80.1 $\mu$m for low speed printing and 265.5, 140.3, and 99.2 $\mu$m for high-speed printing.

\subsection{Case study 3: Validation of the Stability Analysis Method} \label{exp_stability}

\begin{figure}[!t]
\centering
\includegraphics[width=\columnwidth]{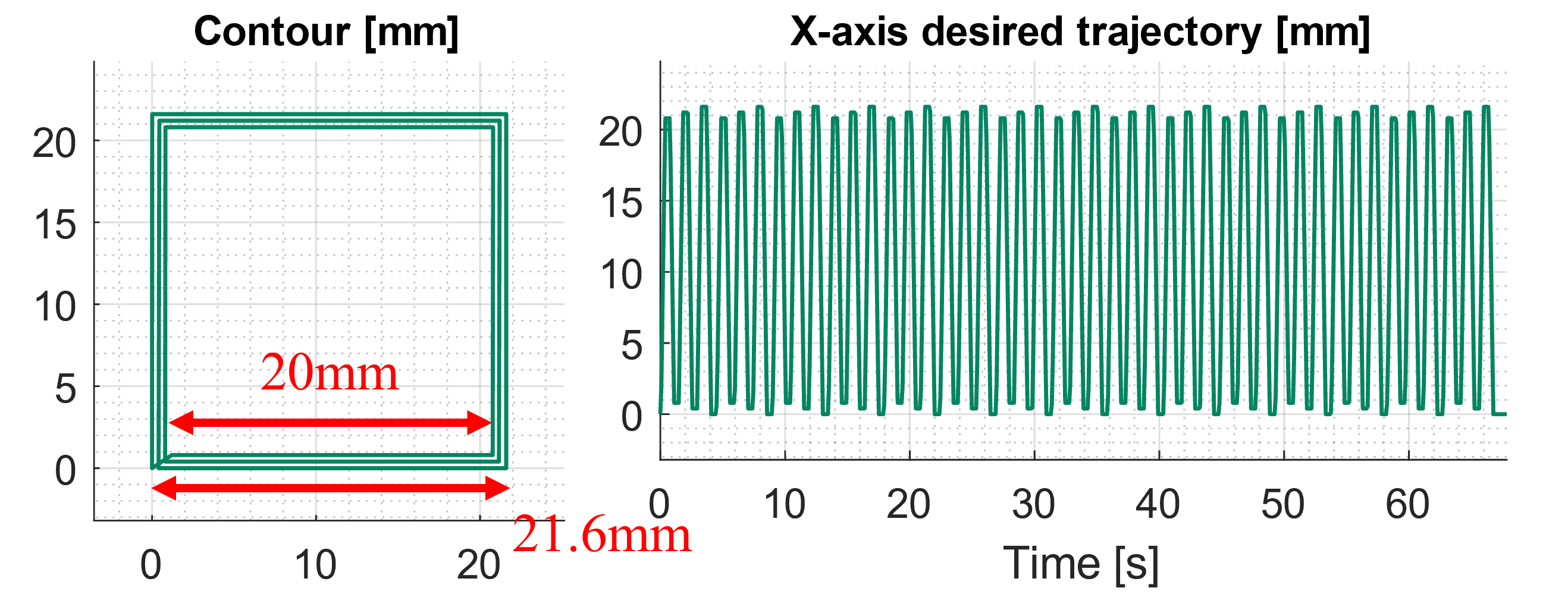}
\caption{The desired shape and the $x$-axis trajectory (the axis containing instability) for Case Study 3.}
\label{Exp3_traj}
\end{figure}

\begin{figure}[!t]
\centering
\includegraphics[width=\columnwidth]{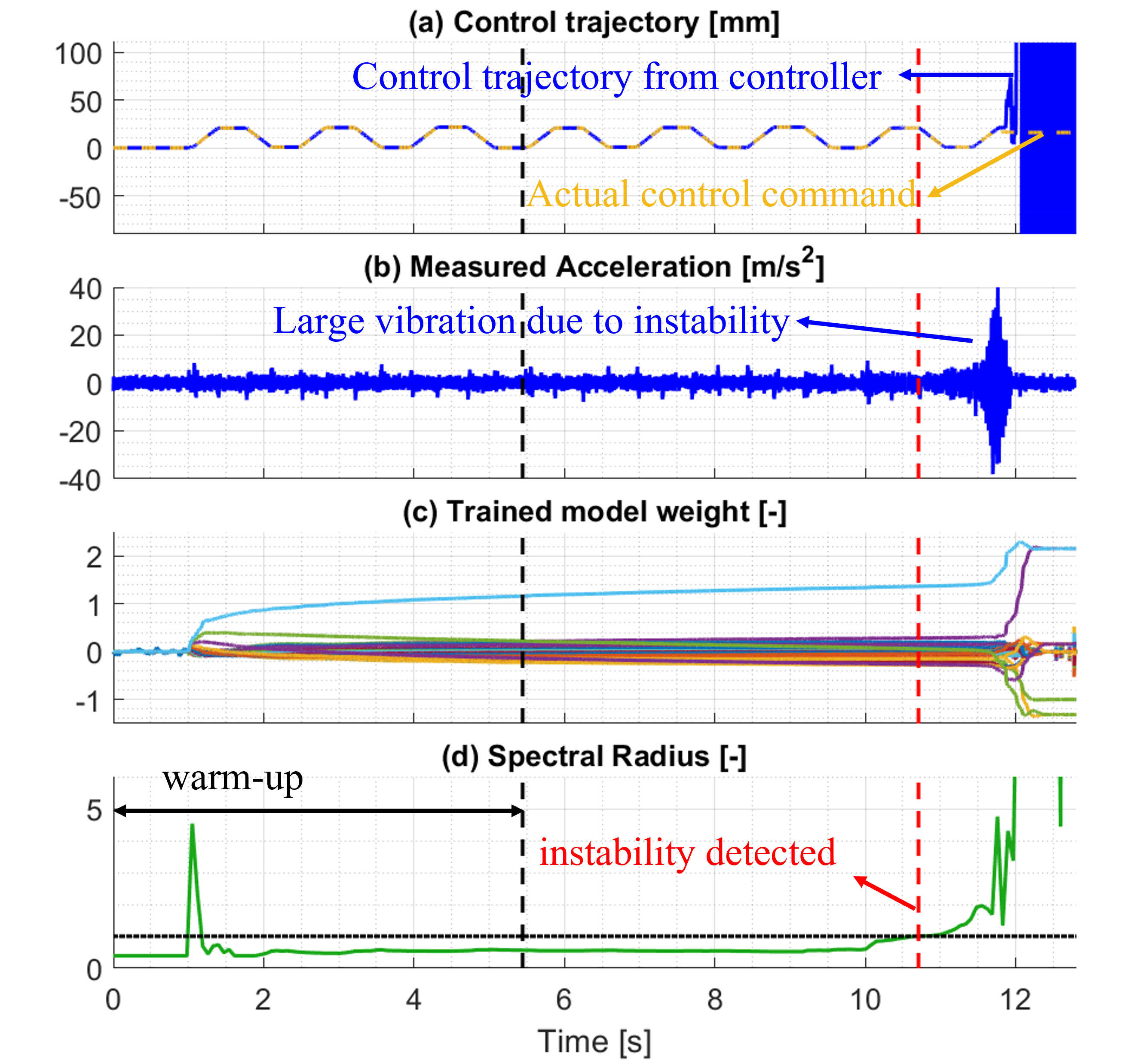}
\caption{The $x$-axis (a) control trajectory, (b) measured acceleration, (c) training results, and (d) stability analysis results for Case Study 3.}
\label{Exp3_results}
\end{figure}

No stability issues were encountered in executing the two case studies reported above. Therefore, in this section, a special case study is created to trigger instability and evaluate the ability of the proposed stability analysis approach to detect the impending instability in advance. Figure \ref{Exp3_traj} shows the motion trajectory used in this case study. It is a hollow square shape that is 3-layer-thick and 15-layer-high. The feed rate, acceleration, and jerk limits used are respectively:
\begin{equation}
    v_{\text{lim}} = 60 \,\text{mm}/\text{s}, \; 
    a_{\text{lim}} = 3 \,\text{m}/\text{s}^2, \; 
    j_{\text{lim}} = 6000 \,\text{m}/\text{s}^3
\end{equation}

Figure~\ref{Exp3_results} shows the control trajectories, measured acceleration, the training results (i.e., the model weights for $\mathbf{G}_h$), and the stability analysis results for this case study. The hybrid FBF controller is warmed up for 5.5~s at the beginning of the motion as it begins to be applied to vibration compensation thereafter. However, at time t~$\approx$~11.5~s, the controller becomes unstable, which outputs diverging control trajectory (due to skipping steps, the actual control command stops) and leads to a sudden spike in acceleration and in some of the trained model weights. Using the stability analysis approach proposed in Section~\ref{stability}, this instability can be detected in advance by the spectral radius approaching unity at t~=~10.7~s, just before the instability sets in. This gives the controller a 0.8~s lead time to take mitigating measures to avert instability. Note that this stability check reported here is performed offline using data gathered online from the $x$-axis accelerometer. However, in practice, it could be implemented online to detect the impending instability and take mitigating actions, e.g., by switching off the hybrid controller or switching the control system to a safe mode that is known to be stable (e.g., the standard FBF controller using a conservative printing speed to ensure the quality).

\section{Conclusion and Future Work} \label{conclusion}

This paper has proposed a physics-guided data-driven controller, called the hybrid filtered basis function (hybrid FBF) controller, for feedforward tracking control of systems with unmodeled dynamics. In contrast with the standard FBF controller that uses only a fixed physics-based model, the hybrid FBF controller consists of a fixed physics-based model and a varying data-driven model that is continuously updated during the execution of the controller to account for unmodeled dynamics. The effect of delays due to data acquisition are incorporated into the formulation of the data-driven portion of the controller to enhance its practicality. Moreover, the hybrid FBF controller has an inherent feedback loop that could lead to instability during its operation, hence unsafe learning. Therefore, a rigorous stability analysis method is proposed to enable the detection and aversion of impending instability.

The performance of the hybrid FBF is demonstrated experimentally using two case studies involving the compensation of the tracking errors of a vibration-prone desktop 3D printer with unmodeled linear and nonlinear dynamics. In both cases, the hybrid FBF controller is shown to significantly outperform the standard FBF controller, particularly in high-speed printing where the effect of unmodeled dynamics are prominent. Moreover, the proposed stability analysis approach is evaluated offline using accelerometer data collected online from the desktop 3D printer. The proposed approach is able to detect impending instability in advance, allowing time for mitigating measures to be taken to avert instability. This bodes well for the practicality of the hybrid FBF controller, as it facilitates the safe implementation of the controller.

There are deficiencies of the hybrid FBF controller that will be addressed in future work. For example, in the second case study involving a retrofitted 3D printer, the hybrid FBF controller had some unresolved quality issues on the printed objects partly due to inconsistent learning from portions of the printed part with different geometries and vibration behaviors. This deficiency could be solved by creating multiple data-driven models for different geometric features and operating conditions. Another opportunity for future work is to explore learning from one system to another, e.g., across a network of 3D printers. This would likely involve the use of privacy-preserving federated learning \cite{kontar2021internet}.

\section*{Acknowledgments}
This work is supported by National Science Foundation grant \#1931950 and a grant from CISCO Systems Inc. A company founded by C.E. Okwudire holds a commercial license for the filtered B spline (FBS) algorithm, which is the version of the standard FBF controller used in this paper.

\bibliographystyle{IEEEtran}
\bibliography{reference}

\vspace{11pt}

\vspace{-33pt}
\begin{IEEEbiographynophoto}{Cheng-Hao Chou}
is a Ph.D. student in the Department of Mechanical Engineering at the University of Michigan. He received the B.S.E. degree in mechanical engineering from National Taiwan University and the M.S.E. degree in mechanical engineering from the University of Michigan. His current research focuses on data-driven methods for control of manufacturing machines.
\end{IEEEbiographynophoto}

\vspace{11pt}

\vspace{-33pt}
\begin{IEEEbiographynophoto}{Molong Duan}
received his B.S. degree from Peking University in 2012 and his M.S. and Ph.D. degrees from the Mechanical Engineering department at the University of Michigan in 2013 and 2018. He is currently an assistant professor at the Department of Mechanical and Aerospace Engineering at the Hong Kong University of Science and Technology. His research interests are the modeling and control of advanced manufacturing systems, flexible structures, and robots. 
\end{IEEEbiographynophoto}

\vspace{11pt}

\vspace{-33pt}
\begin{IEEEbiographynophoto}{Chinedum E. Okwudire}
received his Ph.D. degree in Mechanical Engineering from the University of British Columbia in 2009 and joined the Mechanical Engineering faculty at the University of Michigan in 2011. Prior to joining Michigan, he was the mechatronic systems optimization team leader at DMG Mori USA, based in Davis, CA. His research is focused on exploiting knowledge at the intersection of machine design, control and computing to boost the performance of manufacturing automation systems at low cost. Chinedum has received a number of awards including the CAREER Award from the National Science Foundation; the Young Investigator Award from the International Symposium on Flexible Automation; the Outstanding Young Manufacturing Engineer Award from the Society of Manufacturing Engineers; the Ralph Teetor Educational Award from SAE International; and the Russell Severance Springer Visiting Professorship from UC Berkeley. He has co-authored a number of best paper award winning papers in the areas of control and mechatronics.
\end{IEEEbiographynophoto}

\vfill

\end{document}